\RequirePackage{fix-cm}
\documentclass[
]{rsproca}

\usepackage[T1]{fontenc}
\usepackage[utf8x]{inputenc}
\usepackage[english]{babel}
\usepackage{graphicx}
\usepackage[compress, numbers]{natbib}
\usepackage[absolute]{textpos}

\usepackage{amssymb, amsmath, amsfonts, xspace, units, tabularx}

\newcommand{\numdp}{\ensuremath{N}\xspace}

\newcommand{\condprob}[2]{\ensuremath{p\left(#1\! \mid\! #2 \right)}\xspace}
\newcommand{\joinprob}[2]{\ensuremath{p\left(#1, #2 \right)}\xspace}
\newcommand{\sysX}{\ensuremath{X}\xspace}
\newcommand{\sysY}{\ensuremath{Y}\xspace}

\newcommand{\sysXY}{\ensuremath{{\scriptscriptstyle \sysX \to \sysY}}\xspace}
\newcommand{\sysYX}{\ensuremath{{\scriptscriptstyle \sysY \to \sysX}}\xspace}

\newcommand{\obsX}{\ensuremath{x}\xspace}
\newcommand{\obsY}{\ensuremath{y}\xspace}
\newcommand{\stateX}[1]{\ensuremath{\obsX_{#1}}\xspace}
\newcommand{\stateY}[1]{\ensuremath{\obsY_{#1}}\xspace}

\newcommand{\symbDelay}{\ensuremath{l}\xspace}
\newcommand{\symbDim}{\ensuremath{m}\xspace}
\newcommand{\symbRank}[1]{\ensuremath{k_{i#1}}\xspace}

\newcommand{\symbX}{\ensuremath{\hat \obsX}\xspace}
\newcommand{\symbY}{\ensuremath{\hat \obsY}\xspace}

\newcommand{\segment}{\ensuremath{w}\xspace}
\newcommand{\segnum}{\ensuremath{\eta}\xspace}
\newcommand{\numsegs}{\ensuremath{\numdp_\segnum}\xspace}
\newcommand{\segnumdp}{\ensuremath{\numdp_w}\xspace}

\newcommand{\STE}{\ensuremath{\operatornamewithlimits{\hat T}}\xspace}
\newcommand{\GAMMA}{\ensuremath{\operatornamewithlimits{\hat \gamma}}\xspace}
\newcommand{\barSTE}{\ensuremath{\langle\STE\rangle}\xspace}
\newcommand{\barGAMMA}{\ensuremath{\langle\GAMMA\rangle}\xspace}

\newcommand {\catf}{\ensuremath{\mathfrak{f}}\xspace}
\newcommand {\catn}{\ensuremath{\mathfrak{n}}\xspace}
\newcommand {\cato}{\ensuremath{\mathfrak{o}}\xspace}

\newcommand{\reffig}[1]{Fig.~\ref{#1}}
\newcommand{\reftab}[1]{Tab.~\ref{#1}}

\begin{document}

\title{Assessing directionality and strength of coupling through symbolic analysis: an application to epilepsy patients}

\author{
Klaus Lehnertz$^{1, 2, 3}$ and Henning Dickten$^{1, 2, 3}$}

\address{$^{1}$Department of Epileptology, University of Bonn, Sigmund-Freud-Stra{\ss}e~25, 53105~Bonn, Germany\\
    $^{2}$Helmholtz Institute for Radiation and Nuclear Physics, University of Bonn, Nussallee~14--16, 53115~Bonn, Germany\\
    $^{3}$Interdisciplinary Center for Complex Systems, University of Bonn, Br{\"u}hler Stra{\ss}e~7, 53175~Bonn, Germany}


\corres{Klaus Lehnertz\\
\email{klaus.lehnertz@ukb.uni-bonn.de}}

\begin{abstract}
Inferring strength and direction of interactions from electroencephalographic (EEG) recordings
is of crucial importance to improve our understanding of dynamical interdependencies underlying various physiologic and pathophysiologic conditions in the human epileptic brain.
We here use approaches from symbolic analysis to investigate---in a time-resolved manner---weighted and directed, short- to long-ranged interactions between various brain regions constituting the epileptic network. 
Our observations point to complex spatial-temporal interdependencies underlying the epileptic process and their role in the generation of epileptic seizures, despite the massive reduction of the complex information content of multi-day, multi-channel EEG recordings through symbolisation. 
We discuss limitations and potential future improvements of this approach. 
\end{abstract}

\begin{fmtext}
\end{fmtext}
\maketitle

\begin{textblock*}{20cm}(1cm,25cm)
    Published as Phil. Trans. R. Soc. A \textbf{373}, 2034 (2015). Copyright 2014 the Author(s) Published by the Royal Society.
\end{textblock*}

\section{Introduction}

Epilepsy is one of the most common neurological disorders, second only to stroke, that affects approximately \unit[1]{\%} of the world's population~\cite{Duncan2006, Guerrini2006}. 
In about \unit[25]{\%} of patients, epileptic seizures---the cardinal symptom of epilepsy---can not be controlled by any available therapies~\cite{Schuele2008, Spencer2008, Schramm2008, Kwan2011,Perucca2011}.
In order to enhance the quality of life for epilepsy patients, there is thus a great need for improved therapeutic possibilities.
An epileptic seizure is defined as ``a transient occurrence of signs and/or symptoms due to
abnormal, excessive or synchronous neuronal activity in the brain''~\cite{Fisher2005,Engel2006}.
Epileptic seizures can be divided into two main categories: focal seizures, which appear to originate from a circumscribed brain region (the so called seizure onset zone (SOZ)~\cite{Rosenow2001}) and which may or may not remain restricted to this region, and generalized onset seizures, which appear to engage almost the entire brain.
These concepts of focal and generalized seizures, however, are being challenged by increasing evidence of seizure onset within a network of brain regions (the so called epileptic network) which led to a new approach to classification of seizures and epilepsies~\cite{Berg2011, Lehnertz2014}.
In an epileptic network, all its constituents can contribute to the generation, maintenance, spread, and termination of seizures as well as to the many pathophysiologic phenomena seen during the seizure-free interval~\cite{Bertram1998, Bragin2000, Bartolomei2001, Avoli2002, Spencer2002}.
The concept of an epileptic network may also explain the apparent contradictory findings of seizure precursors that are not confined to the SOZ or its immediate surroundings but can be observed in remote or even contralateral brain regions~\cite{Dalessandro2005, Kalitzin2005, LeVanQuyen2005, Mormann2005, Navarro2005, Federico2005, Meier2007, Feldwisch2011, Aarabi2012, Seyal2014}.

An improved understanding of the epileptic network and its complex dynamics can be achieved with time series analysis techniques that aim at characterising interactions and their properties, namely strength and direction~\cite{Arnhold1999, Osterhage2007, Osterhage2007b, Osterhage2008, Prusseit2008a, Wendling2009, Andrzejak2011, Andrzejak2011b}. 
Over the last years, a number of such bivariate analysis techniques have been proposed, ranging from linear to nonlinear ones.
Comprehensive overviews concerning these techniques and their applications in diverse fields can be found in~\cite{Pikovsky2001, Boccaletti2002, Pereda2005, Gourevitch2006, Marwan2007, Lehnertz2009b, Friedrich2011, Lehnertz2011}. 
Here, we concentrate on information theoretic approaches~\cite{Hlavackova2007}, and particularly on those that utilize symbol sequences~\cite{Hao1989, Bandt2002, Daw2003} derived from empirical time series to characterise interactions.
For the strength of interaction we will consider the order parameter \GAMMA proposed by Liu~\cite{Liu2004} and for the direction of interaction the symbolic transfer entropy~\cite{Staniek2008, Staniek2009}.
Both approaches allow for a robust and computationally fast quantification of the strength and the preferred direction of information flow between time series from observed data. 

Previous studies~\cite{Staniek2008, Staniek2009} that investigated directed interactions in the human epileptic brain provided evidence that the aforementioned approaches allow one to reliably identify the hemisphere containing the SOZ without observing actual seizure activity. 
In these studies, however, analyses were restricted to data from patients with mesial temporal lobe epilepsies and to recordings from within the mesial temporal lobes. 
Here, we extend these studies in several aspects.
First, we investigate multi-day (on average about 4 days), multi-channel electroencephalographic data recorded invasively from 11 patients that suffered from seizures with different anatomical onset locations.
Second, we investigate---in a time-resolved manner---changes in interdependencies between all sampled brain regions, thus taking into account interactions for a variety of physiologic and pathophysiologic conditions.  
Third, we estimate \emph{both} strength and direction of interactions in order to effectively distinguish the various coupling regimes (uncoupled, weak to strong couplings) that may be observable in the human epileptic brain and in order to avoid misinterpretations~\cite{Osterhage2008, Staniek2009}. 
Fourth, we address the question of who is driving preferentially whom in large-scale epileptic brain networks and whether consistent changes in directed interactions can be identified when approaching epileptic seizures.

\section{Strength and direction of interaction from symbolic analysis}
In the following, we denote with $\stateX{i}:=x(i\Delta t)$ and $\stateY{i}:=y(i\Delta t)$, $i=1, \ldots, \numdp$, time series of observables of systems \sysX and \sysY recorded simultaneously with sampling interval $\Delta t$. 
We make use of the seminal work of Bandt and Pompe~\cite{Bandt2002} and derive symbol sequences from reordering the amplitude values of time series. 
Let \symbDelay and \symbDim denote the embedding delay and embedding dimension, which were chosen appropriately for symbolisation.
Then \symbDim amplitude values
\mbox{$
    s_i = \left\{ \stateX{i}, \stateX{i+\symbDelay}, \ldots, \stateX{i + \symbDelay(\symbDim - 1)} \right\}
$}
for a given, but arbitrary timestep $i$ are arranged in ascending order
\mbox{$
    \left\{ \stateX{i + \symbDelay(\symbRank{1} - 1)} \leq \stateX{i + \symbDelay(\symbRank{2} - 1)} \leq \ldots \leq \stateX{i + \symbDelay(\symbRank{m} - 1)}\right\}
$}
with rank \symbRank{i}.
In case of equal amplitude values the rearrangement is carried out according to the associated index $k$, i.e., for $\stateX(i+(k_{i1}-1)l)=\stateX(i+(k_{i2}-1)l)$ we write $\stateX(i+(k_{i1}-1)l)\leq \stateX(i+(k_{i2}-1)l)$ if $k_{i1}<k_{i2}$. This ensures that every $s_i$ is uniquely mapped onto one of the $\symbDim!$ possible permutations, and a permutation symbol is defined as
\begin{equation}
    \symbX_i := \left(\symbRank{1}, \symbRank{2}, \ldots, \symbRank{m} \right).
    \label{eq:symbolization}
\end{equation}
The symbol sequence $\symbY_i$ for system \sysY is defined in complete analogy.

A robust method to estimate the strength of interaction from symbol sequences was proposed by Liu~\cite{Liu2004}.
It is based on consistent changing tendencies of temporal permutation entropies
$H_\sysX(\segment_\segnum) = \sum_i p(\symbX_i) \log p(\symbX_i)$,
where the sum runs over all possible symbols.
$H_\sysY(\segment_\segnum)$ is defined in complete analogy.
In order to estimate these changing tendencies, symbol sequences $\symbX_i$ and $\symbY_i$, $i = 1, \ldots, \numdp$ are split into \numsegs segments $\segment_\segnum$ with $\segnum = 1, \ldots, \numsegs$, consisting of \segnumdp datapoints, and permutation entropies $H_\sysX$ and $H_\sysY$ are estimated for each segment $\segment_\segnum$.
$H_\sysX(\segment_\segnum)$ and $H_\sysY(\segment_\segnum)$ can be expected to be similar if some functional relationship---in the sense of generalized synchronization~\cite{Rulkov1995}---exists between states of \sysX and \sysY.
The changing tendency for a time series from system \sysX can be quantified with
\begin{equation}
    S_\sysX(\segment_\segnum) = 
                \begin{cases}
                    1   & \text{if\quad} H_\sysX(\segment_{\segnum + 1}) > H_\sysX(\segment_\segnum)\\
                    -1  & \text{else,}
                \end{cases}
\end{equation}
and in complete analogy for a time series from system $S_\sysY$. A measure for the strength of interaction between systems \sysX and \sysY can then be defined as
\begin{equation}
    \GAMMA = \frac{1}{\numsegs} \sum_{\segnum = 1}^{\numsegs} S_\sysX (\segment_\segnum) S_\sysY(\segment_\segnum).
    \label{eq:gamma}
\end{equation}
This index will be around zero for independent time series and close to unity for generalized synchronization. 
Note that \GAMMA might attain slightly negative values.

Symbolic transfer entropy (STE)~\cite{Staniek2008, Staniek2009} has been proposed as a computational fast and robust method to quantify the direction of interaction between coupled systems.
It is based on Schreiber's transfer entropy~\cite{Schreiber2000} that allows one to distinguish effectively driving and responding elements and to detect asymmetry in the interaction of systems.
STE overcomes some of the limitations of previous techniques that aim at estimating transfer entropy from time series, such as requiring fine tuning of parameters and being highly sensitive to noise contributions~\cite{Kaiser2002, Verdes2005, Lungarella2007}.
More specifically, STE makes use of relative frequencies of symbols to estimate  the transition probabilities required to quantify the deviation from the generalised Markov property,
$\condprob{\stateX{i}}{\stateX{i-1}, \stateY{i-1}} = \condprob{\stateX{i}}{\stateX{i-1}}$, where \condprob{\cdot}{\cdot} denotes the conditional transition probability density. 
With transfer entropy this deviation is formulated as a Kullback-Leibler entropy between \condprob{\stateX{i}}{\stateX{i-1}, \stateY{i-1}} and \condprob{\stateX{i}}{\stateX{i-1}}.

With given symbol sequences $\symbX_i$ and $\symbY_i$, symbolic transfer entropy is then defined as
\begin{equation}
 \STE_\sysYX = \sum \joinprob{\symbX_i}{\symbX_{i-1}, \symbY_{i-1}} \log\frac{\condprob{\symbX_i}{\symbX_{i-1}, \symbY_{i-1}}}
          {\condprob{\symbX_i}{\symbX_{i-1}}},
  \label{eq:STE}
\end{equation}
where the sum runs over all symbols.
$\STE_\sysXY$ is defined in complete analogy, and the directionality index $\STE:= \STE_\sysXY - \STE_\sysYX$ allows one to quantify the preferred direction of information flow between systems \sysX and \sysY. 
\STE attains positive values for unidirectional coupling with
\sysX as the driver, negative values for \sysY driving \sysX, and $\STE \approx 0$ for symmetric bidirectional coupling.

Since its invention, STE has been used to study interactions in various disciplines ranging from quantum~\cite{Kowalski2010}, plasma~\cite{Melzer2014}, and laser physics~\cite{Nian-Qiang2012} via neurology~\cite{Blain-Moraes2013}, cardiology~\cite{Jun2012} and anesthesiology~\cite{Ku2011,Jordan2013,Lee2013,Untergehrer2014} to the neurosciences~\cite{Martini2011}.
In addition, extensions have been proposed that allow one to estimate the dominating direction of information flow even from short or transient signals~\cite{Martini2011}. A recent modification, termed partial symbolic transfer entropy~\cite{Kugiumtzis2013b, Papana2013},
extends STE for multivariate time series and can help distinguish between direct and indirect interactions.

Estimating interaction properties with information-theoretic measures is prone to biases and statistical errors which depend on the method used and on the characteristics of the data~\cite{Hlavackova2007,Hahs2011,Barnett2012,Smirnov2013,Haruna2013}. 
For both, the order parameter \GAMMA and the directionality index \STE, asymptotic distributions are unknown, and in the following we apply averaging procedures (both in time and space) to decrease the variance of estimation error and consider both aspects of interaction, strength and direction, to minimise the risk for misinterpretations and to effectively distinguish the various coupling regimes.

\begin{figure}
    \center
    \includegraphics[scale=0.75]{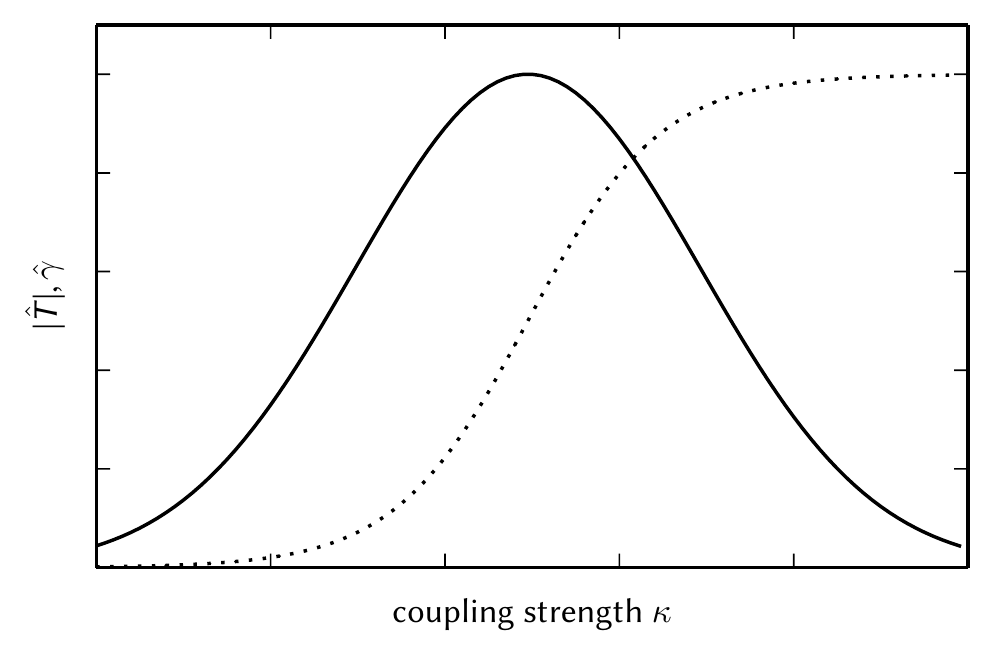}
    \caption{Schematic dependence of the directionality index (\STE, solid line) and of the order parameter (\GAMMA, dotted line) on the coupling strength in some driven-responder system.}
    \label{fig:indexscheme}
\end{figure}

Based on previous investigations that employed bivariate time series analysis techniques to characterise \emph{both} strength and direction of interactions in studies on coupled model systems~\cite{Palus2007, Waddell2007, Osterhage2008, Staniek2008, Staniek2009}, we expect the order parameter \GAMMA and the directionality index \STE to depend on the coupling strength $\kappa$ in some driver--responder system as follows (cf.~\reffig{fig:indexscheme}): 
\GAMMA increases monotonously with increasing $\kappa$ until $\GAMMA=1$ for fully synchronised systems. 
This dependency thus correctly reflects an increasing strength of interaction due to the increased coupling.
On the other hand, \STE correctly detects the direction of interaction for some intermediate values of $\kappa$ only.
Moreover, $\STE \approx 0$ for uncoupled ($\kappa=0$ and $\GAMMA=0$) or for fully synchronised systems ($\GAMMA=1$). 
Distinguishing these coupling regimes thus requires checking the strength of interaction. 
When investigating interacting systems using empirical data, the coupling strength is usually not known a priori, but we expect the aforementioned dependencies to hold (at least approximately) for such data.

\section{Measuring strength and direction of interactions in the human epileptic brain}
\begin{figure}
    \center
    \includegraphics[width=0.75\textwidth]{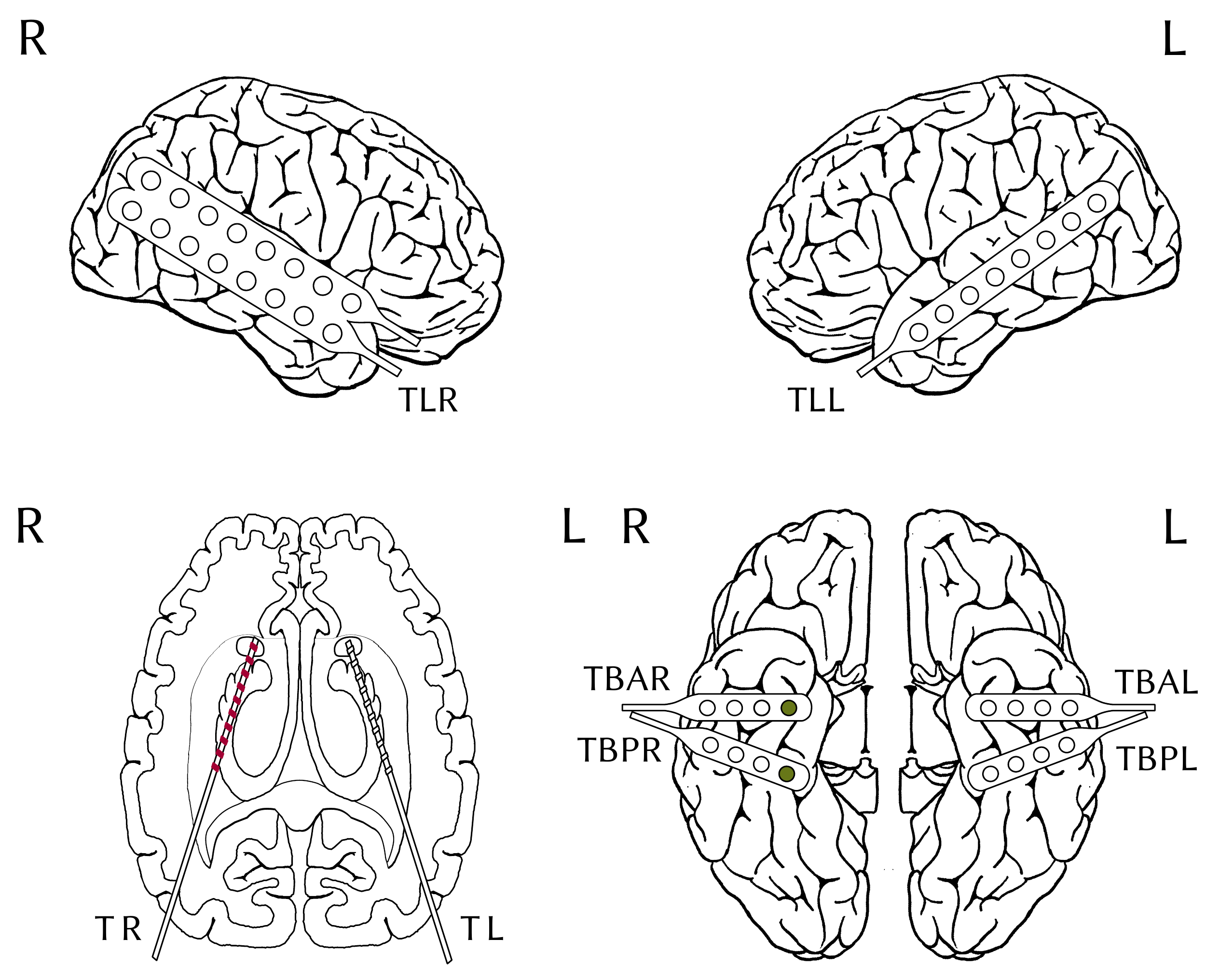}
    \caption{Schematics of implanted electrodes from a patient with seizures originating from the right mesial temporal lobe. 
Intrahippocampal depth electrodes were implanted stereotaxically along the longitudinal axis of the hippocampal formation from an occipital approach with the amygdala as the target for the most anterior electrode. 
Each catheter-like, \unit[1]{mm} thick silastic electrode contained 10 cylindrical contacts of a nickel--chromium alloy (\unit[2.5]{mm}) every \unit[4]{mm}. 
Subdural strip electrodes consisted of 4, 8, or 16 stainless steel contacts with a diameter of \unit[2.2]{mm}, embedded in a silastic (intercontact spacing of \unit[10]{mm}). 
These electrodes were inserted through burr holes and were placed over the inferior and lateral temporal cortex. 
Colours indicate location categories to which electrode contacts belong: \catf, red; \catn, green; \cato, white.}
    \label{fig:implscheme}
\end{figure}

We retrospectively analysed strength and direction of interactions between various brain regions by applying the aforementioned methods 
to multi-channel, multi-day electroencephalographic (EEG) data recorded intracranially in 11 patients (4 women, 7 men; mean age at onset of epilepsy 10.5 years, range 0--32 years; mean duration of epilepsy 28 years, range 5--53 years). 
These data had been recorded during the presurgical evaluation of intractable focal seizures with different anatomical onset locations. 
The patients had signed informed consent that their clinical data might be used and published for research purposes.
The study protocol had previously been approved by the ethics committee of the University of Bonn.
EEG data were recorded from, on average, 48 sites (range 21--73; cf.~\reffig{fig:implscheme}), band-pass-filtered between \unit[1--45]{Hz} and sampled at \unit[200]{Hz} (sampling interval $\Delta t = $ \unit[5]{ms}) using a 16~bit analog-to-digital converter. 
Recordings lasted, on average, \unit[98.6]{h} (range \unit[25.4--206.2]{h}) during which seven seizures/patient (range 1--21) were captured.
In five patients, seizures originated from the mesial temporal lobe, in two patients from the lateral temporal lobe, and in four patients from the frontal lobe.
Postoperatively, all patients were seizure-free~\cite{Engel1993b}.

As a compromise between the statistical accuracy for the calculation of interaction indices and approximate stationarity~\cite{Blanco1995}, we divided the data into a sequence of non-overlapping segments of \unit[20.48]{s} duration (corresponding to 4096 data points).
This allowed us to calculate \GAMMA and \STE for each combination of pairs of recording sites in a time-resolved manner.
If not stated otherwise, all calculations were performed with \symbDim = 5, \symbDelay = 3, \segnumdp = 2048, and \numsegs = 204~\cite{Cao2004, Staniek2007}.

Using knowledge concerning location and extent of the SOZ, which is defined by the electrode contacts showing initial seizure activity~\cite{Rosenow2001}, we assigned all electrode contacts to three location categories:
\begin{description}
    \item[focal (\catf):] electrode contacts located within the SOZ (on average for all patients~\unit[13.0]{\%} of all contacts, varying between~\unit[5.0]{\%} and~\unit[20.8]{\%});
    \item[neighbour (\catn):] electrode contacts not more than two contacts distant to those from category \catf (on average~\unit[9.1]{\%}, varying between~\unit[2.6]{\%} and~\unit[14.9]{\%});
    \item[other (\cato):] all remaining electrode contacts (on average~\unit[77.9]{\%}, varying between~\unit[66.2]{\%} and~\unit[87.7]{\%}). 
\end{description}

This allowed us to concisely analyze interactions between brain regions and to perform inter-subject comparisons, despite the high variability in location and extent of the individual SOZ, which necessitated an electrode implantation tailored to the individual patient.

Since the order parameter \GAMMA is a symmetric and the directionality index \STE an antisymmetric measure, we will present our findings on the strength resp.\ direction of interactions from six combination categories (\catf-\catf, \catf-\catn, \catf-\cato, \catn-\catn, \catn-\cato, \cato-\cato; the direction of interaction is encoded in the sign of \STE).

\begin{figure}
    \center
    \includegraphics[scale=0.75]{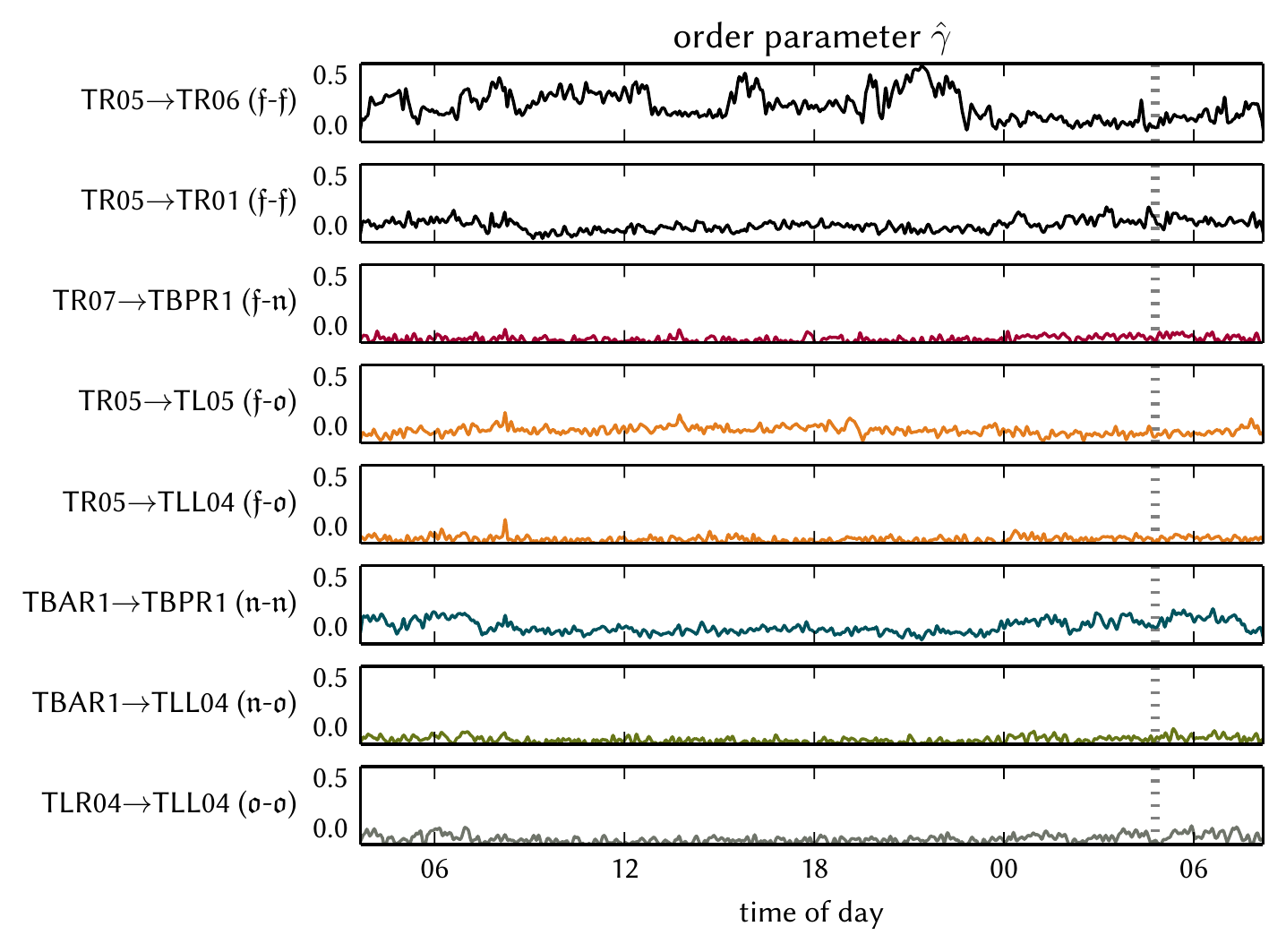}
    \includegraphics[scale=0.75]{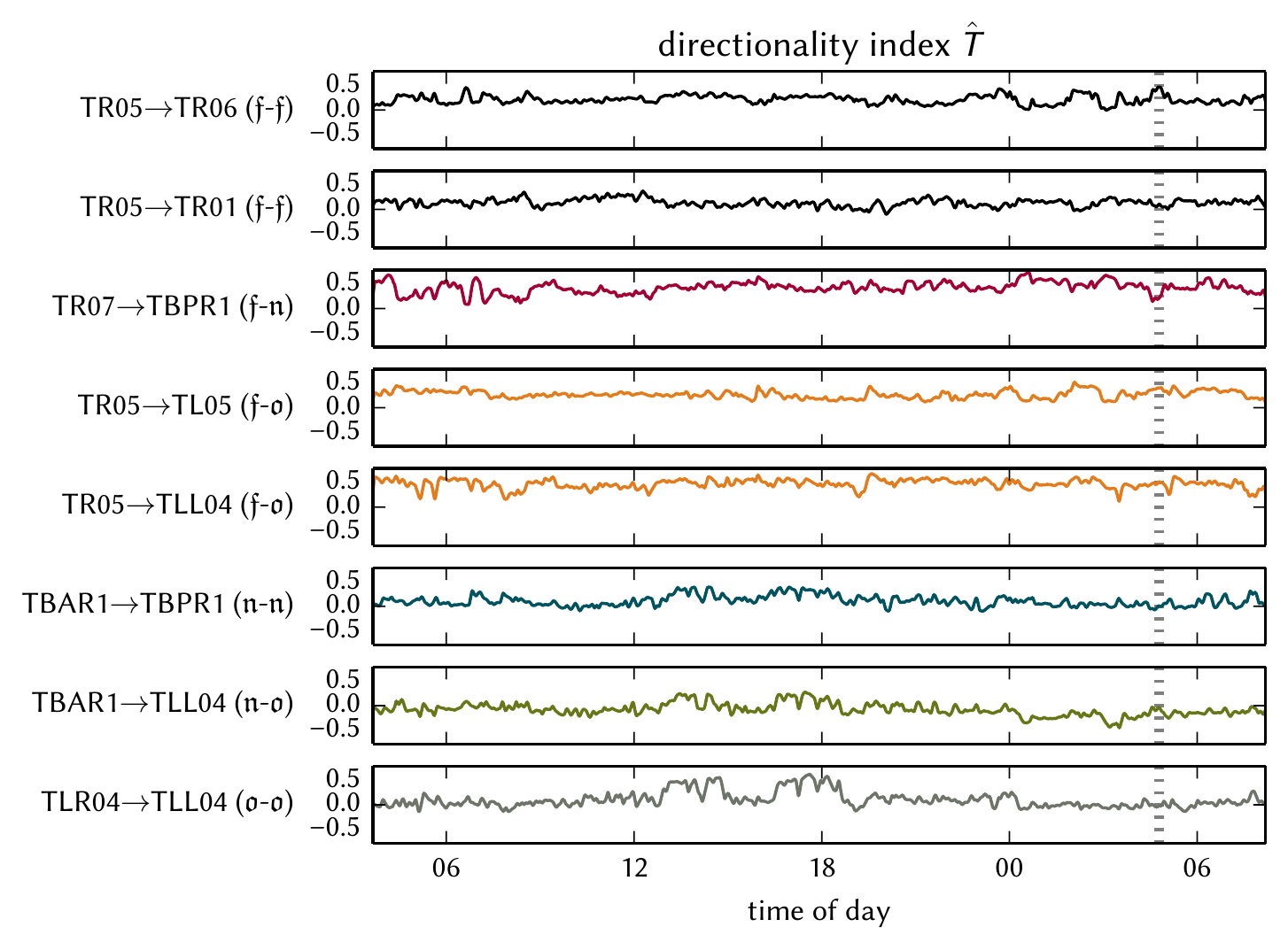}
    \caption{Exemplary temporal evolutions of the order parameter \GAMMA (upper plot) and the directionality index \STE (lower plot) for interactions between various brain regions (from top to bottom): within the SOZ (\catf-\catf), between SOZ and its neighbourhood (\catf-\catn), between SOZ and other brain areas (\catf-\cato), neighbour--neighbour interactions (\catn-\catn), as well as neighbour--other (\catn-\cato) and other--other interactions (\cato-\cato). 
The dashed vertical line indicates onset of an epileptic seizure that lasted for approx.~\unit[100]{s}.
For readability, time profiles are smoothed using a Hamming window (\unit[10]{min} duration). Time 00 indicates midnight.}
    \label{fig:res_temporal_evolution}
\end{figure}

In~\reffig{fig:res_temporal_evolution} we show, as an example, temporal evolutions of the strength (\GAMMA) and the direction (\STE) of interactions between various brain regions over a time course of \unit[28]{h}, during which a seizure was captured. The data were calculated from intracranial EEG recordings of the patient whose electrode implantation scheme is shown in~\reffig{fig:implscheme}.
For only few pairs of recording sites can we observe temporarily increased values of \GAMMA. 
These increases do not appear to coincide with seizure-related activities, but are nevertheless confined to interactions within the SOZ (\catf-\catf) and to neighbour--neighbour interactions (\catn-\catn).
Recognisable indications for directed interactions can be mainly observed for combination categories involving the SOZ, and particularly for  interactions between the SOZ and its neighbourhood as well as with other, remote brain regions. 
These observations indicate the SOZ to be a driving brain region (cf. Refs.~\cite{Palus2001a,Osterhage2007b,Osterhage2008,Osterhage2008a}) with interactions that are not only local but involve large parts of the epileptic network.

\begin{figure}
    \center
    \includegraphics[scale=0.75]{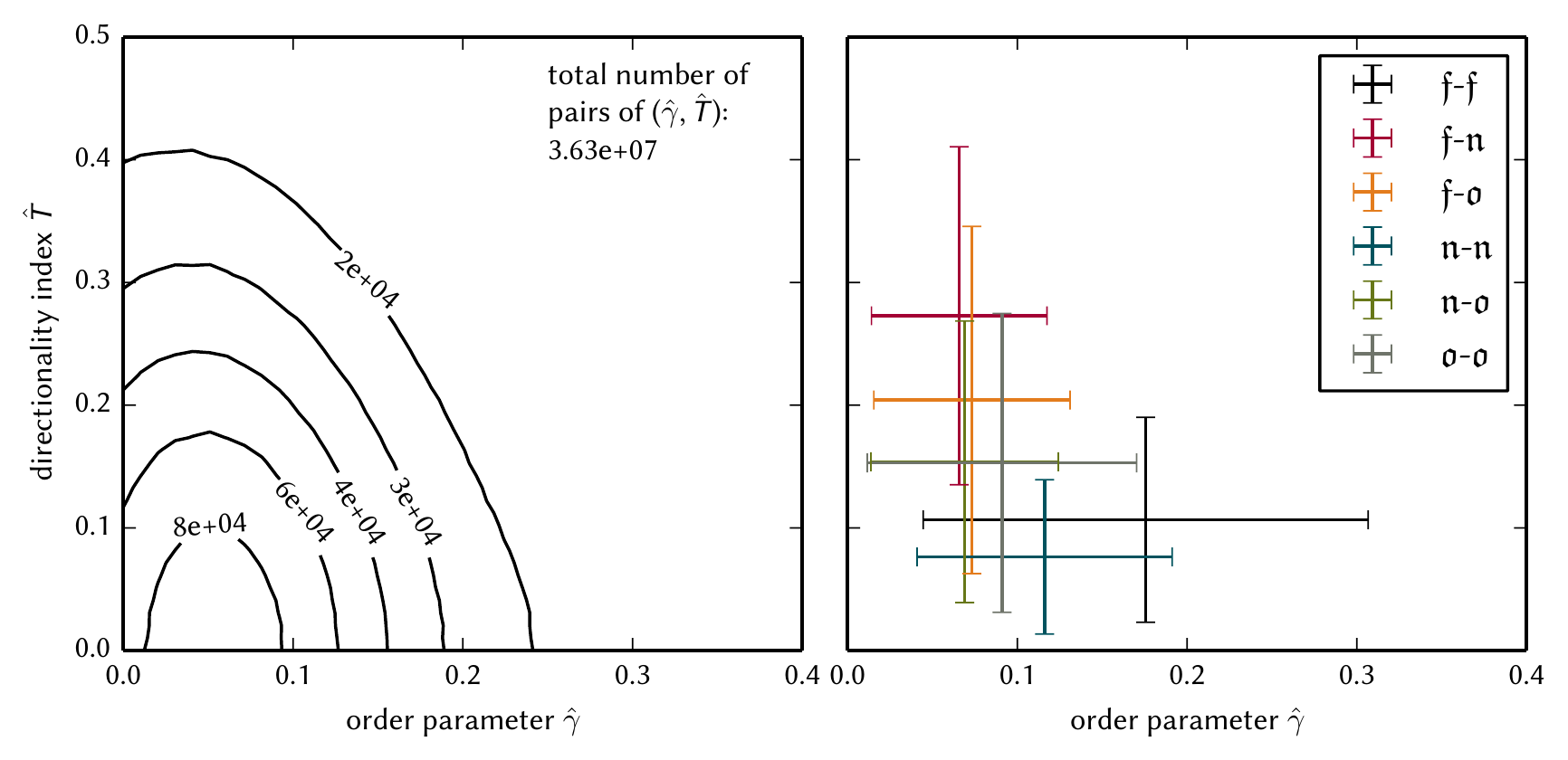}
    \caption{Left: Contour lines of equal frequency in a bi-dimensional histogram of more than $3.6 \cdot 10^7$ pairs of values of the order parameter \GAMMA and of the directionality index \STE ($\STE \geq 0$) estimated from intracranial EEG data recorded from one patient  (cf.~\reffig{fig:implscheme}) over nine days for all combinations of electrode contacts (seizure-free interval only, i.e. data from a presumed pre-ictal phase (\unit[4]{h} duration), the ictal, and the post-ictal phase (\unit[30]{min} duration) of four seizures were excluded). Right: Temporal and spatial means and standard deviations of \GAMMA and of the absolute value of \STE for all categories of combinations of recordings sites.} 
    \label{fig:res_direction_vs_strength_01341}
\end{figure}

From the data shown in~\reffig{fig:res_temporal_evolution}, it can be deduced that, for the most part, indications for directed interactions go along with low to intermediate strengths of interactions, while high values of the latter are accompanied by indications for symmetric bidirectional couplings. 
Our findings for all pairs of recording sites (see~\reffig{fig:res_direction_vs_strength_01341} left)
and for all
combination categories (see~\reffig{fig:res_direction_vs_strength_01341} right) support our expectations mentioned above (see~\reffig{fig:indexscheme}) and underline the importance of investigating both strength and direction of interaction in order to 
avoid misinterpretations.

\begin{figure}
    \center
    \includegraphics[scale=0.75]{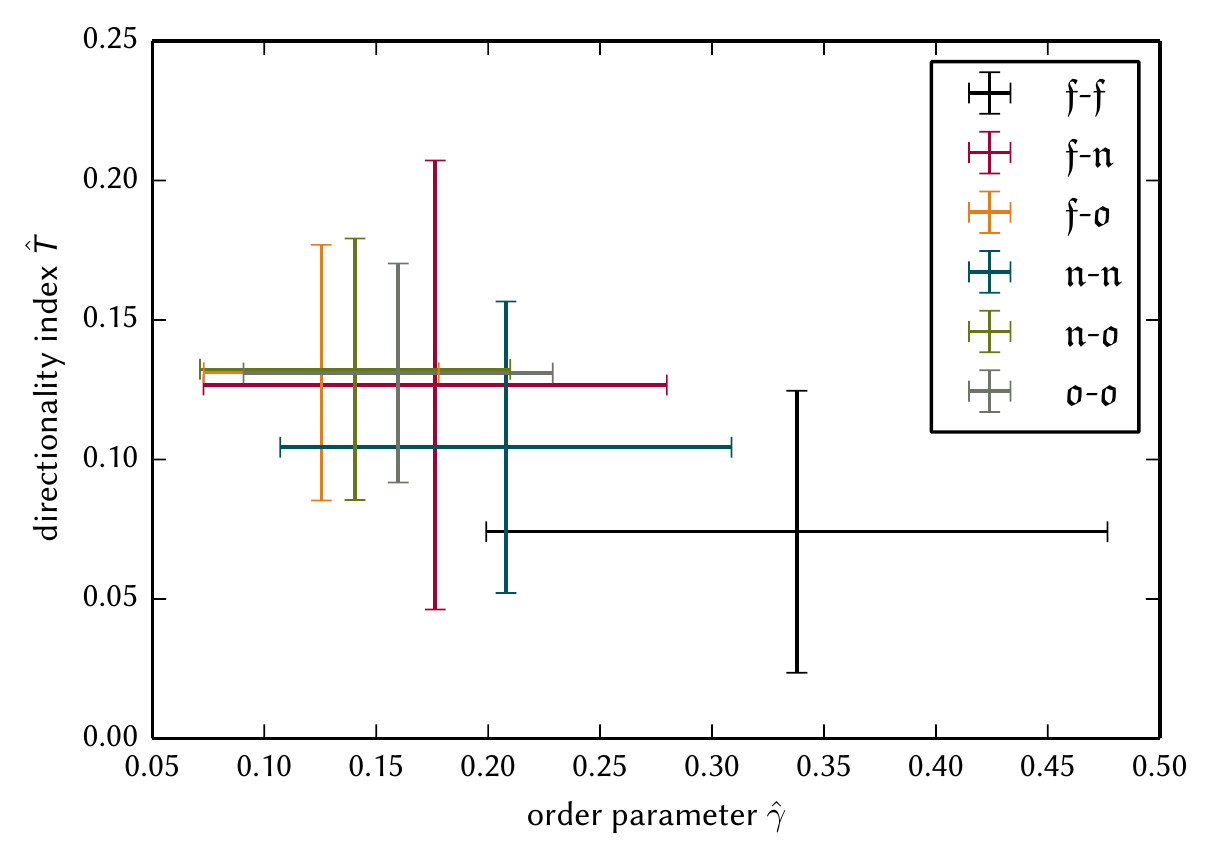}
    \caption{Means and standard deviations of patient-wise means of \GAMMA and of the absolute value of \STE for all categories of combinations of recordings sites for the 11 patients.}
    \label{fig:res_direction_vs_strength_all}
\end{figure}

We now investigate whether the aforementioned observations of strongest interactions confined to the SOZ and/or its immediate surroundings extend beyond exemplary data.
In~\reffig{fig:res_direction_vs_strength_all}, we show \STE and \GAMMA for all categories of combinations of recordings sites estimated from the data from all patients.
On this group level, strengths of interactions appear to decrease with an increasing distance from the SOZ and its surroundings, which is in line with a number of previous studies~\cite{BenJacob2007a,Lai2007,Osterhage2007,Schevon2007,Ortega2008,Zaveri2009,Warren2010,Andrzejak2011,Bettus2011,Ortega2011,Palmigiano2012}.
If we search, however, for the maximum strength of interactions among all combinations of recording sites in each patient separately, we observe this maximum (and the next 10 lower values) for interactions far off the SOZ (category \cato-\cato) in 9 of 11 patients. 
In one patient, categories \catf-\catf and \cato-\cato rank almost equally among the ten highest values.
In another patient, categories \catf-\catf and \catn-\catn clearly rank highest among all combinations of recording sites.
These observations  on a single patient level are in contrast to many of the previous studies mentioned before, and it remains to be shown whether differences can be related to the much lower number of recording sites taken into account before or to possibly different sensitivities of applied estimators for the strength of interactions (see, e.g.,~\cite{Ansari2006,Osterhage2007,Lehnertz2009b}).

\begin{figure}
    \center
    \includegraphics[scale=0.7]{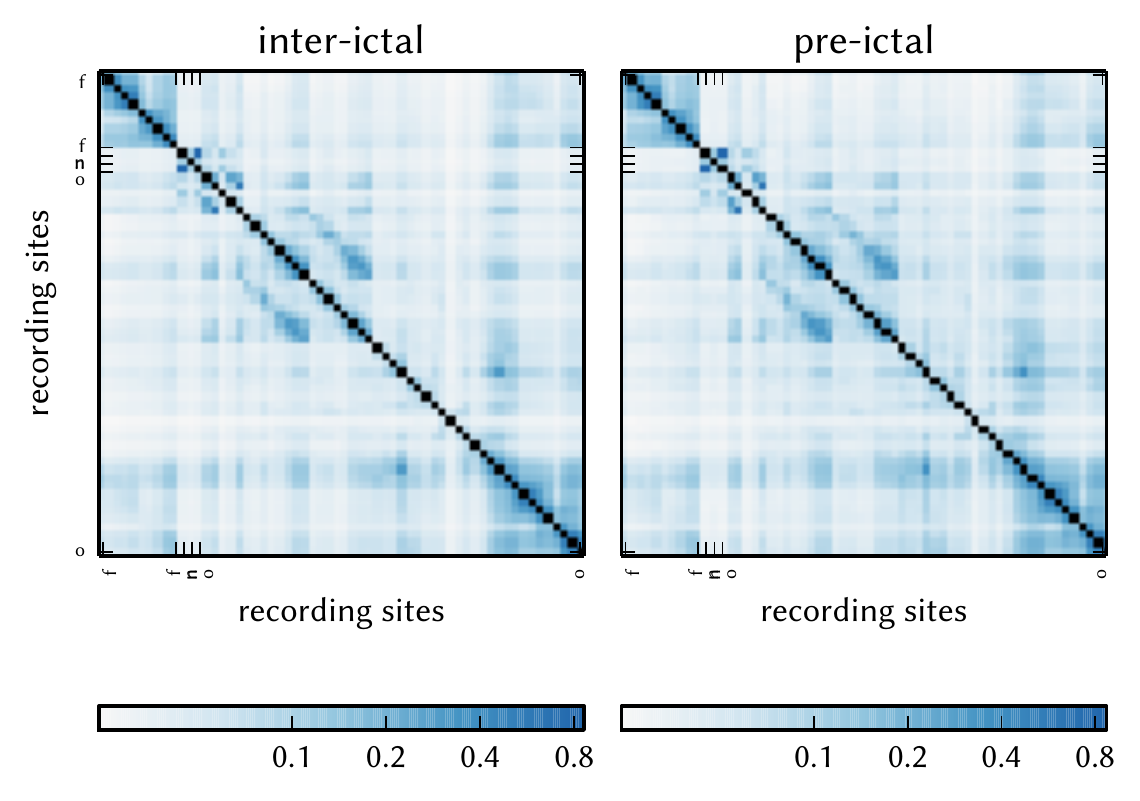}
    \includegraphics[scale=0.7]{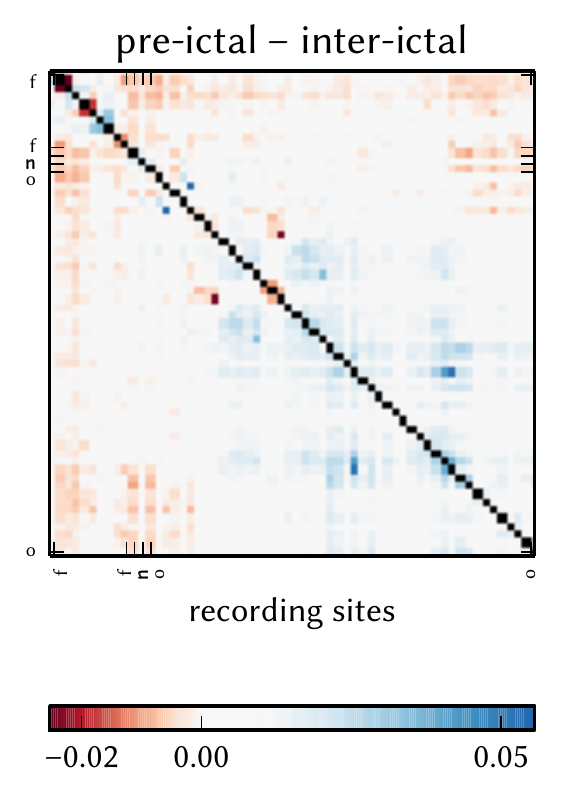}\\
    \includegraphics[scale=0.7]{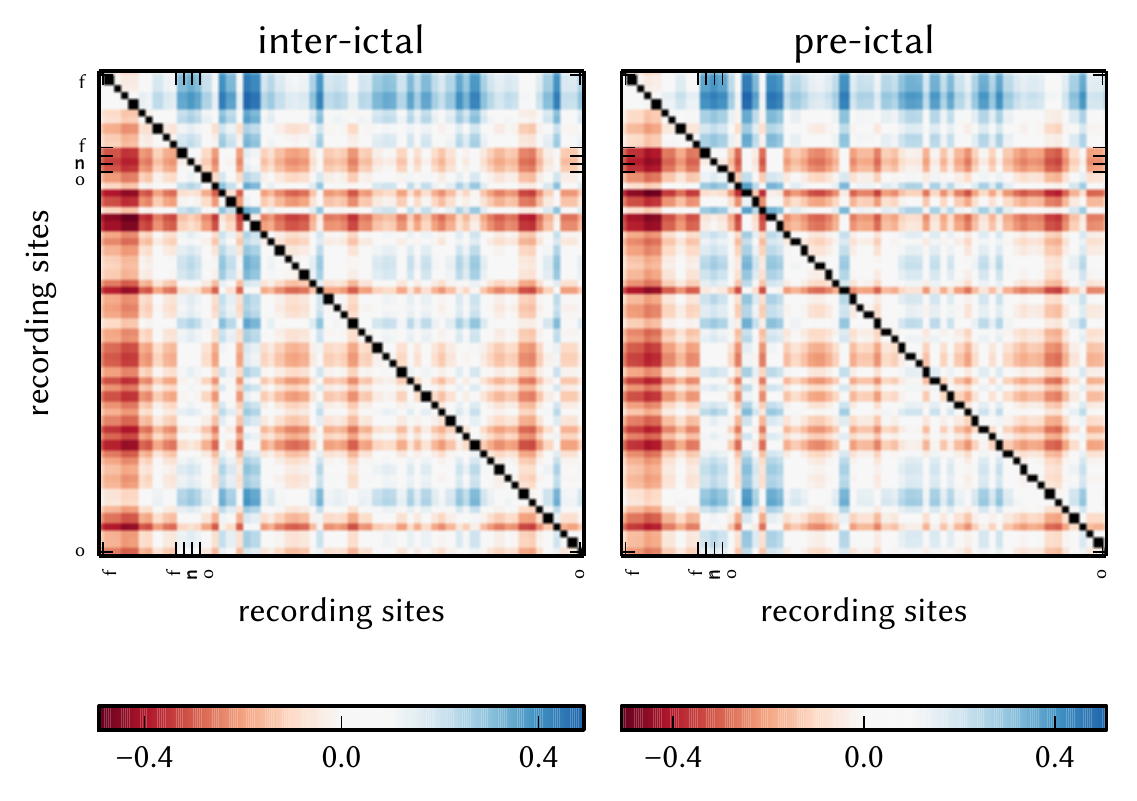}
    \includegraphics[scale=0.7]{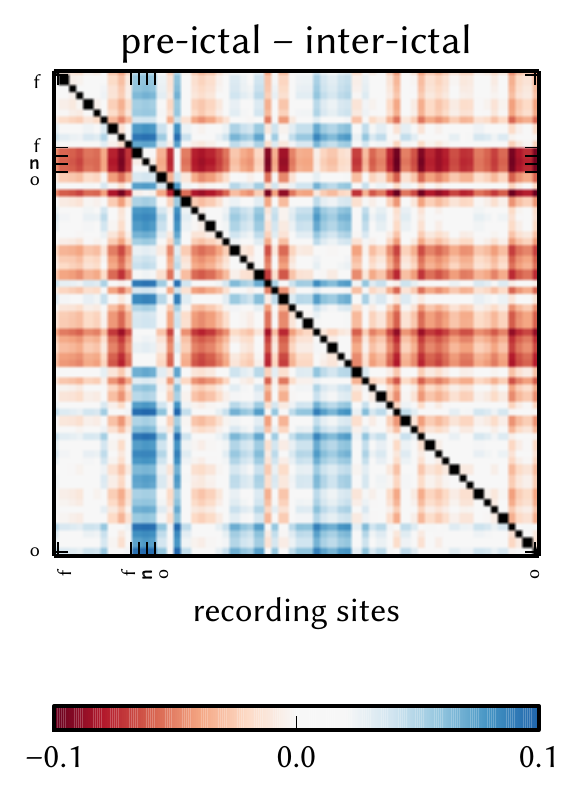}
    \caption{Averaged strengths (top) and directions (bottom) of all pair-wise interactions during inter-ictal (left) and pre-ictal (middle) periods estimated from an intracranial EEG recording (cf.~\reffig{fig:implscheme}) that lasted nine days and during which four seizures were captured. 
    Positive (negative) values of \STE indicate that sites listed on the ordinate drive (are being driven by) sites listed on the abscissa. Differences between pre-ictal and inter-ictal data are shown in the right column.} 
    \label{fig:res_pre_inter}
\end{figure}

So far we only considered data from the seizure-free interval (inter-ictal state) for our analyses. 
We now proceed by addressing the question whether changes in directed interactions between brain regions constituting the epileptic network can be identified prior to seizures (pre-ictal state). 
Our previous investigations~\cite{Lehnertz2011b} indicate that directed interactions between the SOZ and homologous contralateral sites decrease (i.e. from a unidirectional driving to a symmetric bidirectional coupling) during the pre-ictal states in 11 out of 15 patients with mesial temporal lobe epilepsy.
In this study, we also employed symbolic transfer entropy but restricted our analyses to intrahippocampal recordings.
In the following, we will again assume that a pre-ictal phase of \unit[4]{h} duration exists~\cite{Mormann2005} and compare the distributions of values of \STE from the pre-ictal periods with those from inter-ictal periods (all data that were recorded at least \unit[4]{h} prior to and \unit[30]{min} after a seizure) and taking into account all interactions.
In~\reffig{fig:res_pre_inter}, we show exemplary matrices representing the spatial distribution of averaged strengths and directions of all pair-wise interactions during pre-ictal and inter-ictal periods.
For both periods, highest values of \GAMMA can be observed 
for interactions within the SOZ (category \catf-\catf; right hippocampal formation; sites TR01--TR10) and its surrounding brain areas (category \catn-\catn; sites TBAR1--TBAR3 and TBPR1--TBPR3) but also for interactions in the ipsilateral neocortex, particularly at parieto-occipital sites (category \cato-\cato; TLR06--TLR08 and TLR14--TLR16). 
Strongest interactions also involve the most anterior contact in the contralateral hippocampal formation (category \cato-\cato; particularly site TL01) and parieto-occipital sites in the contralateral neocortex (category \cato-\cato; sites TLL07--TLL08).
In line with~\cite{Mormann2008b}, we observe---in both hippocampal formations---a gap in the inter-regional strength of interactions namely within the entorhinal cortex and within the hippocampus, that indicate the existence of independent rhythms (mostly in the frequency band \unit[0.5--7]{Hz}) in different subregions of the human medial temporal lobe, produced by autonomous generators.
Prominent differences between data from the pre-ictal and the inter-ictal periods are visible for the SOZ, its neighbourhood, homologous sites within the contralateral mesial temporal lobe, and for interactions between inferior and lateral sites from the contralateral temporal cortex.
We observe both, a pre-ictal increase and decrease of \GAMMA, in line with previous studies that used other, mostly phase-based estimators for the strength of interactions~\cite{Mormann2000,Mormann2003a,LeVanQuyen2005,Winterhalder2006c,Kuhlmann2010,Feldwisch2011,Wang2011b,Zheng2014}.

If we assume that the direction of interactions can best be resolved at intermediate strengths (see~\reffig{fig:indexscheme}), we accordingly observe high absolute values of \STE particularly for interactions involving categories \catf-\catn, \catf-\cato, and \catn-\cato. 
During both the pre-ictal and the inter-ictal periods, the SOZ (right hippocampal formation; sites TR01--TR10) appears to persistently drive all other brain regions, while these regions appear to drive structures near the SOZ (sites TBAR1--TBAR3 and TBPR1--TBPR3).      
Nevertheless, we also observe directed interactions between brain regions that are apparently not associated with the SOZ (category \cato-\cato).
As a result, no clear-cut directionality patterns can be observed from the differences between data from the pre-ictal and the inter-ictal periods, although strongest changes in directionality, i.e. more pronounced indications for a unidirectional driving coupling, are confined to brain regions surrounding the SOZ (category \catn-\catn; sites TBAR1--TBAR3 and TBPR1--TBPR3).
\begin{figure}
    \center
    \includegraphics[scale=0.75]{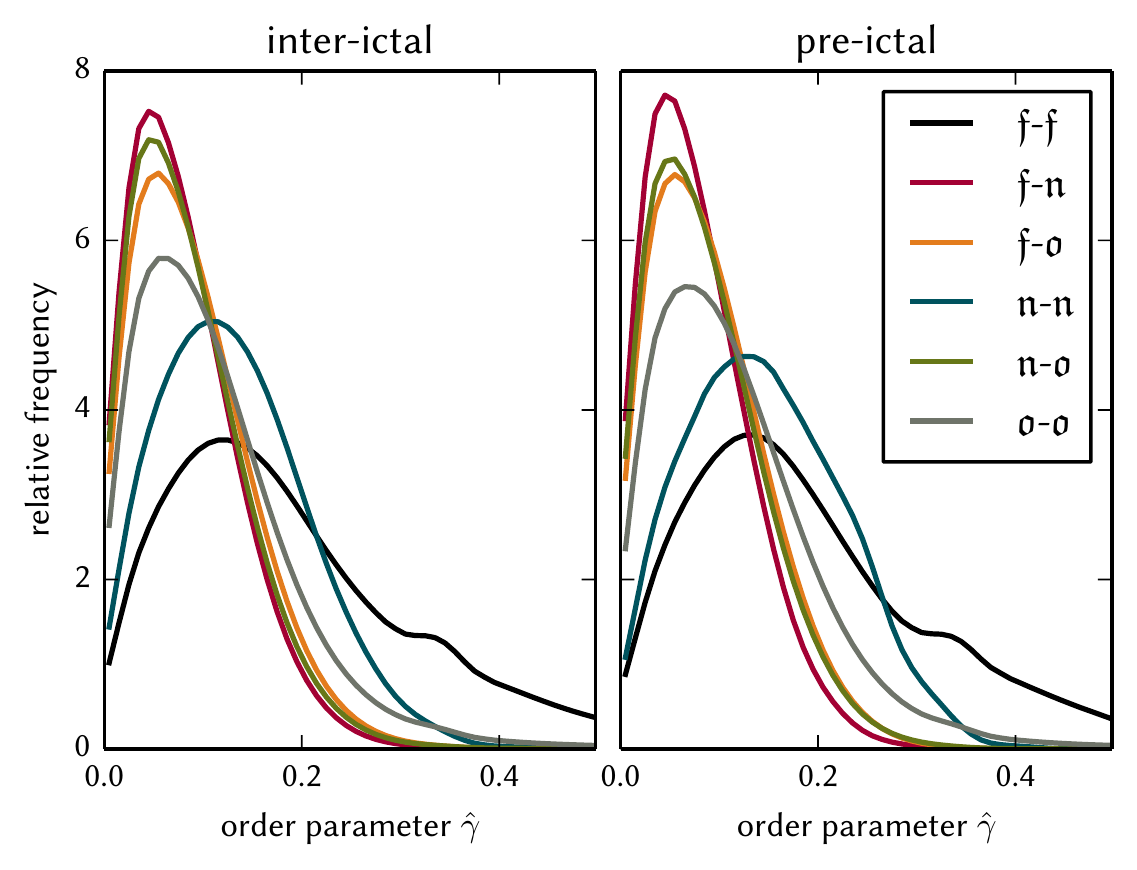}
    \includegraphics[scale=0.75]{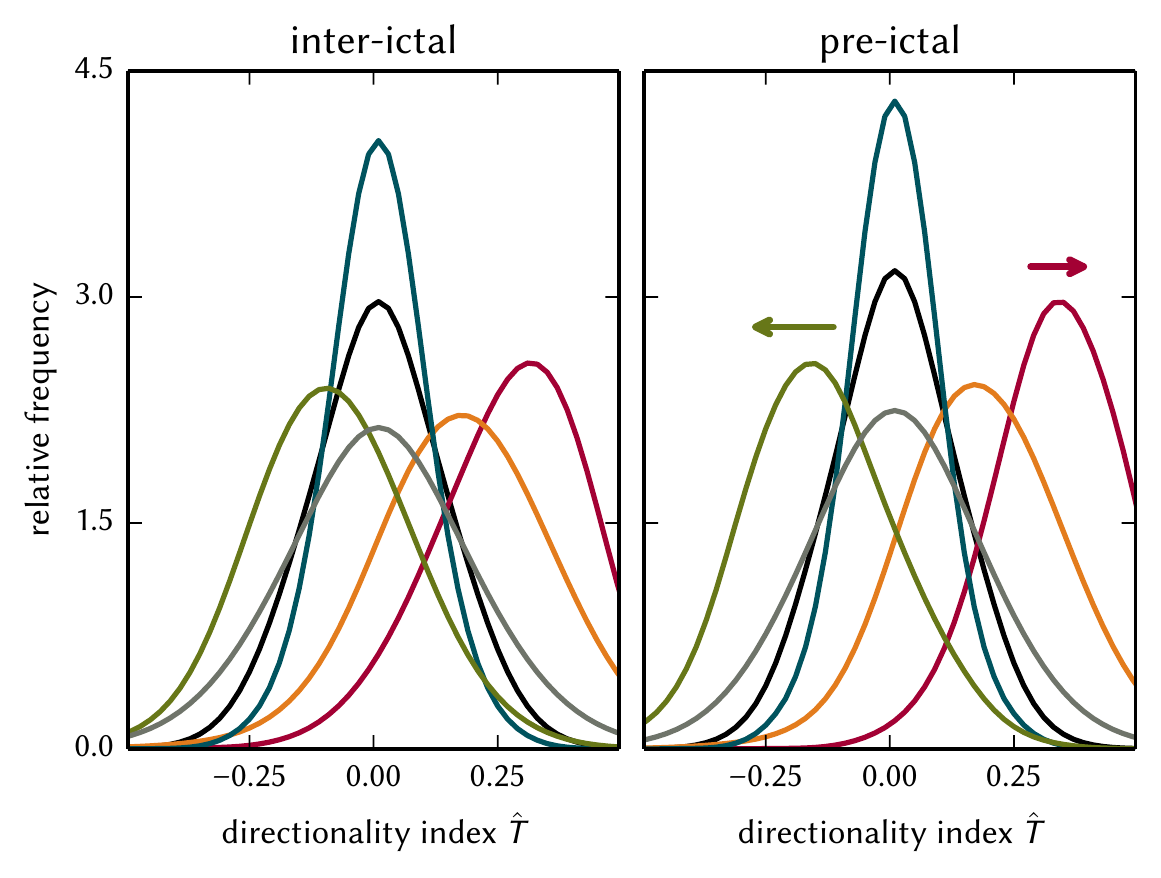}
    \caption{Frequency distributions of order parameter \GAMMA (top) and directionality index \STE (bottom) for all categories of combinations of recordings sites estimated from intracranial EEG data (cf.~\reffig{fig:implscheme}) recorded over nine days. Left: data from inter-ictal periods. Right: data from pre-ictal periods.}
    \label{fig:res_GAMMA_TE_hist}
\end{figure}
In~\reffig{fig:res_GAMMA_TE_hist}, we show the inter-ictal and pre-ictal frequency distributions of indices \GAMMA and \STE for all categories of combinations of recordings sites. 
Pronounced relative changes (> 20 \%) can only be observed for directed interactions between brain regions surrounding the SOZ and all other sampled regions (category \catn-\cato; inter-ictal: $-0.09\pm0.18$; pre-ictal: $-0.15\pm0.16$) and for directed interactions between the SOZ and its neighbourhood (category \catf-\catn; inter-ictal: $0.26\pm0.15$; pre-ictal: $0.33\pm0.13$). 

In order to investigate whether the aforementioned findings extend beyond exemplary data, we estimated---for each combination category (but excluding those that involve the same location category)---the number of patients for which pre-ictally an increase or a decrease (of at least \unit[5]{\%}), or no change of the order parameter and of the directionality index can be observed as compared to the inter-ictal periods (see~\reftab{tab:table1}).
\begin{table}
    \center
    \begin{tabularx}{\textwidth}{l*6{>{\centering\arraybackslash}X}@{}}
            \hline
						{interactions between} & \multicolumn{3}{c}{order parameter \GAMMA} &\multicolumn{3}{c}{directionality index \STE}\\
            brain regions & decrease & increase & else & decrease & increase & else\\
            \hline
            SOZ $\rightleftarrows$ neighbour   & 3 & 6 & 2 & 2 & 5 & 4 \\
            SOZ $\rightleftarrows$ other      & 1 & 5 & 5 & 5 & 3 & 3 \\
            neighbour $\rightleftarrows$ other & 0 & 4 & 7 & 3 & 5 & 3 \\
        \hline
    \end{tabularx}
    \caption{Number of patients for which an increase, a decrease, or no change (else; 	
    changes in amplitude values lower than \unit[5]{\%} or undefined relative changes) of 
    \GAMMA and of \STE can be observed pre-ictally as compared to the inter-ictal periods.} 
    \label{tab:table1}
\end{table}
For about half the cases, the strength of interactions between the SOZ, its surrounding and other brain regions (categories \catf-\catn and \catf-\cato) increased during the pre-ictal periods. The strength of interactions between brain regions not involving the SOZ (category \catn-\cato) remained unchanged pre-ictally in the majority of cases. 
The direction of interactions increased pre-ictally in about half the cases for interdependencies between the SOZ and its surrounding (category \catf-\catn) as well as between brain regions not involving the SOZ (category \catn-\cato). The direction of interactions decreased pre-ictally in about half the cases for interdependencies between the SOZ and other brain regions (category \catf-\cato).

In~\reftab{tab:res:pre_vs_inter}, we report the number of patients for which \STE indicates a preferred direction of interactions (preferentially driving vs. preferentially responding) during pre-ictal and inter-ictal periods. For this purpose we calculated, for each investigated combination category, the temporal and spatial median of \STE and of \GAMMA (denoted as \barSTE and \barGAMMA, respectively), 
and regarded the interaction as preferential driving (responding) if \barSTE was positive (negative) and the accompanying values of \barGAMMA were from the interquartile range of the respective distribution of values of \GAMMA,
We could neither observe a clear-cut indication for a preferential driving nor for a preferential responding between the investigated brain regions.
Pre-ictally, the preferential driving of surrounding brain region by the SOZ (category \catf-\catn; as seen inter-ictally) appears to be lost, however, the driving of surrounding brain regions by other regions (category \catn-\cato) can be observed more often. 
Although speculative, these observation probably point to an effective \emph{inhibitory surround}~\cite{Prince1967,Elger1983} that prevents the spreading of epileptiform activities during the seizure-free interval.
\begin{table}
    \center
    \begin{tabularx}{\textwidth}{l*4{>{\centering\arraybackslash}X}@{}}
    \hline
    {interactions between} &  \multicolumn{2}{c}{driving ($A \rightarrow B$)}  &  \multicolumn{2}{c}{responding ($A \leftarrow B$)}\\
    brain regions $A\rightleftarrows B$& inter-ictal & pre-ictal & inter-ictal & pre-ictal\\
    \hline
    SOZ $\rightleftarrows$ neighbour   & 3 & 0 & 2 & 2\\
    SOZ $\rightleftarrows$ other      & 2 & 3 & 1 & 3\\
    neighbour $\rightleftarrows$ other & 2 & 4 & 4 & 5\\
    \hline
    \end{tabularx}
    \caption{Number of patients for which \STE (using \GAMMA as selection criterion) indicates
     interactions between different brain regions as preferentially driving or responding during
     inter-ictal and pre-ictal periods.}
    \label{tab:res:pre_vs_inter}
\end{table}

\section{Conclusion}
Using approaches from symbolic analysis, we investigated---in a time-resolved manner---changes in strength and direction of short- to long-ranged interactions between brain regions constituting the epileptic network.
We here employed the order parameter \GAMMA and the directionality index \STE (derived from symbolic transfer entropy), both allowing for a robust and computationally fast quantification of the mentioned interaction properties.
We followed previous recommendations~\cite{Osterhage2008, Staniek2009} to investigate both aspects of interactions in order to avoid misinterpretations and to allow an effective differentiation between various coupling regimes.

Analysing multi-day, multi-channel EEG recordings from 11 epilepsy patients, we could confirm some previous findings concerning the strength of interactions that had been derived with frequency-based, phase-based, state-space-based, or statistical approaches. 
Summarising these observations, the strength of interactions decreases with an increasing distance from the seizure-onset zone (SOZ) and its surroundings during the seizure-free (inter-ictal) periods, while it exhibits a possibly location-specific increase or decrease during the pre-ictal periods. 
Our findings obtained from investigating interdependencies between all sampled brain regions, however, indicate that the maximum strength of interactions can be observed for interactions between brain regions far off the SOZ in the majority of cases.  
It thus remains to be shown whether the dynamics of SOZ can indeed be characterised by an elevated local synchrony or elevated strength of interactions~\cite{BenJacob2007a,Lai2007,Schevon2007,Ortega2008,Bettus2011,Ortega2011,Palmigiano2012} and whether this property can help to identify the SOZ in the presurgical evaluation. 
It also remains to be shown whether the location-specific changes of the strength of interactions (increase or decrease) during the pre-ictal periods can be regarded as precursors of epileptic seizures. 
Such an investigation would require using statistical methods for testing the significance of the predictive performance of the applied analysis 	techniques~\cite{Andrzejak2003,Andrzejak2009,Winterhalder2003,Kreuz2004,Schelter2006,Wong2007,Snyder2008,Feldwisch2011}.
The same holds for the direction of interactions since the unambiguous inference of the direction of interactions remains an unsolved problem, despite some developments to test the significance of directionality estimates~\cite{Porta2002, Thiel2006,Romano2009,Faes2010}.

We here considered a spatial average over pre-defined location categories using knowledge concerning location and extent of the SOZ and observed in some patients a preferential driving between some brain regions while in other patients directionality was inverted for the very same brain regions. 
This ambiguity persisted when comparing data from the inter-ictal and the pre-ictal periods.
Due to these inconsistencies we can not yet give a clear-cut answer to the question of who is driving whom in large-scale epileptic brain networks.

There are, by now, only a few studies that investigated directed interaction in the epileptic brain~\cite{Franaszczuk1994,Franaszczuk1998,LeVanQuyen1998,Bartolomei2001,Wendling2001,Chavez2003,Smirnov2005b,Wilke2009,Lehnertz2011b}, and these studies mostly concentrated on seizures or other epileptiform activities (such as epileptic spikes) and/or were mostly restricted to selected recording sites.
A comparison of our findings with those obtained in these studies is thus rather constricted.

The inference of directed interactions from empirical data may be limited by a number of influencing factors.
Bivariate analysis, as used here, is often not sufficient to reveal the correct interaction structure, i.e. distinguishing direct and indirect interactions.
An extension to the concept of symbolic transfer entropy that aims at accounting for the presence of other observed confounding variables has recently been proposed (so called partial symbolic transfer entropy~\cite{Kugiumtzis2013b,Papana2013}), but its 
suitability for characterising directed interactions in epileptic brain networks remains to be investigated.
It is well known that interactions between and within brain regions may be delayed, with time delays reaching up to \unit[200]{ms} depending on region and functions~\cite{Nunez1995}, and not accounting for such delays may lead to spurious indications for directed interactions.
An improved characterisation of such delayed, directed interactions may be achieved with recently proposed analysis techniques that explicitly incorporate time delays in estimates for information transfer~\cite{Frenzel2007,Pompe2011,Runge2012,Runge2012b,Dickten2014}.

In the future, we expect further improvements in the data-driven characterisation of weighted and directed, short- to long-ranged interactions in evolving epileptic brain networks to advance our understanding of the dynamical disease epilepsy, which may guide new developments for individualised diagnosis, treatment, and control.

\section*{Acknowledgment}
The authors are grateful to G. Ansmann, S. Porz, and C. Geier for critical comments on earlier versions of the manuscript.
This work was supported by the Deutsche Forschungsgemeinschaft (Grant No: LE 660/5-2).



\begin{thebibliography}{100}
\expandafter\ifx\csname urlstyle\endcsname\relax
  \providecommand{\doi}[1]{doi:\discretionary{}{}{}#1}\else
  \providecommand{\doi}{doi:\discretionary{}{}{}\begingroup
  \urlstyle{rm}\Url}\fi

\bibitem{Duncan2006}
Duncan JS, Sander JW, Sisodiya SM, Walker MC, 2006 Adult epilepsy.
\newblock \emph{Lancet} \textbf{367}, 1087--1100.
\newblock \doi{10.1016/S0140-6736(06)68477-8}

\bibitem{Guerrini2006}
Guerrini R, 2006 Epilepsy in children.
\newblock \emph{Lancet} \textbf{367}, 499--524.
\newblock \doi{10.1016/S0140-6736(06)68182-8}

\bibitem{Schuele2008}
Schuele S, L\"uders HO, 2008 Intractable epilepsy: management and therapeutic
  alternatives.
\newblock \emph{Lancet Neurol.} \textbf{7}, 514--524.
\newblock \doi{10.1016/S1474-4422(08)70108-X}

\bibitem{Spencer2008}
Spencer S, Huh L, 2008 Outcomes of epilepsy surgery in adults and children.
\newblock \emph{Lancet Neurol.} \textbf{7}, 525--537.
\newblock \doi{10.1016/S1474-4422(08)70109-1}

\bibitem{Schramm2008}
Schramm J, 2008 Temporal lobe epilepsy surgery and the quest for optimal extent
  of resection: {A} review.
\newblock \emph{Epilepsia} \textbf{49}, 1296--1307.
\newblock \doi{10.1111/j.1528-1167.2008.01604.x}

\bibitem{Kwan2011}
Kwan P, Schachter SC, Brodie MJ, 2011 Drug-resistant epilepsy.
\newblock \emph{N. Engl. J. Med.} \textbf{365}, 919--926

\bibitem{Perucca2011}
Perucca E, Tomson T, 2011 The pharmacological treatment of epilepsy in adults.
\newblock \emph{Lancet Neurol.} \textbf{10}, 446--456.
\newblock \doi{10.1016/S1474-4422(11)70047-3}

\bibitem{Fisher2005}
Fisher RS, van Emde~Boas W, Blume W, Elger CE, Genton P, Lee P, {Engel Jr} J,
  2005 Epileptic seizures and epilepsy: definitions proposed by the
  {I}nternational {L}eague {A}gainst {E}pilepsy ({ILAE}) and the
  {I}nternational {B}ureau for {E}pilepsy ({IBE}).
\newblock \emph{Epilepsia} \textbf{46}, 470--472.
\newblock \doi{10.1111/j.0013-9580.2005.66104.x}

\bibitem{Engel2006}
{Engel Jr} J, 2006 Report of the {ILAE} {C}lassification {C}ore {G}roup.
\newblock \emph{Epilepsia} \textbf{47}, 1558--1568.
\newblock \doi{10.1111/j.1528-1167.2006.00215.x}

\bibitem{Rosenow2001}
Rosenow F, L\"uders H, 2001 Presurgical evaluation of epilepsy.
\newblock \emph{Brain} \textbf{124}, 1683--1700.
\newblock \doi{10.1093/brain/124.9.1683}

\bibitem{Berg2011}
Berg AT, Scheffer IE, 2011 New concepts in classification of the epilepsies:
  Entering the 21\textsuperscript{st} century.
\newblock \emph{Epilepsia} \textbf{52}, 1058--1062.
\newblock \doi{10.1111/j.1528-1167.2011.03101.x}

\bibitem{Lehnertz2014}
Lehnertz K, Ansmann G, Bialonski S, Dickten H, Geier C, Porz S, 2014 Evolving
  networks in the human epileptic brain.
\newblock \emph{Physica D} \textbf{267}, 7--15.
\newblock \doi{10.1016/j.physd.2013.06.009}

\bibitem{Bertram1998}
Bertram EH, Zhang DX, Mangan P, Fountain N, Rempe D, 1998 Functional anatomy of
  limbic epilepsy: a proposal for central synchronization of a diffusely
  hyperexcitable network.
\newblock \emph{Epilepsy Res.} \textbf{32}, 194--205.
\newblock \doi{10.1016/S0920-1211(98)00051-5}

\bibitem{Bragin2000}
Bragin A, Wilson CL, {Engel Jr} J, 2000 Chronic epileptogenesis requires
  development of a network of pathologically interconnected neuron clusters: a
  hypothesis.
\newblock \emph{Epilepsia} \textbf{41 (Suppl.~6)}, S144--S152.
\newblock \doi{10.1111/j.1528-1157.2000.tb01573.x}

\bibitem{Bartolomei2001}
Bartolomei F, Wendling F, Bellanger JJ, R\`egis J, Chauvel P, 2001 Neural
  networks involving the medial temporal structures in temporal lobe epilepsy.
\newblock \emph{Clin. Neurophysiol.} \textbf{112}, 1746--1760

\bibitem{Avoli2002}
Avoli M, {D'Antuono} M, Louvel J, K\"ohling R, Biagini G, Pumain R,
  {D'Arcangelo} G, Tancredi V, 2002 Network and pharmacological mechanisms
  leading to epileptiform synchronization in the limbic system in vitro.
\newblock \emph{Prog. Neurobiol.} \textbf{68}, 167--207

\bibitem{Spencer2002}
Spencer SS, 2002 Neural networks in human epilepsy: Evidence of and
  implications for treatment.
\newblock \emph{Epilepsia} \textbf{43}, 219--227.
\newblock \doi{10.1046/j.1528-1157.2002.26901.x}

\bibitem{Dalessandro2005}
D'Alessandro M, Vachtsevanos G, Esteller R, Echauz J, Cranstoun S, Worrell G,
  Parish L, Litt B, 2005 A multi-feature and multi-channel univariate selection
  process for seizure prediction.
\newblock \emph{Clin. Neurophysiol.} \textbf{116}, 506--516

\bibitem{Kalitzin2005}
Kalitzin S, Velis D, Suffczynski P, Parra J, {Lopes da Silva} F, 2005
  Electrical brain-stimulation paradigm for estimating the seizure onset site
  and the time to ictal transition in temporal lobe epilepsy.
\newblock \emph{Clin. Neurophysiol.} \textbf{116}, 718--728.
\newblock \doi{10.1016/j.clinph.2004.08.021}

\bibitem{LeVanQuyen2005}
{Le Van Quyen} M, Soss J, Navarro V, Robertson R, Chavez M, Baulac M,
  Martinerie J, 2005 Preictal state identification by synchronization changes
  in long-term intracranial {EEG} recordings.
\newblock \emph{Clin. Neurophysiol.} \textbf{116}, 559--568

\bibitem{Mormann2005}
Mormann F, Kreuz T, Rieke C, Andrzejak RG, Kraskov A, David P, Elger CE,
  Lehnertz K, 2005 On the predictability of epileptic seizures.
\newblock \emph{Clin. Neurophysiol.} \textbf{116}, 569--587.
\newblock \doi{10.1016/j.clinph.2004.08.025}

\bibitem{Navarro2005}
Navarro V, Martinerie J, {Le Van Quyen} M, Baulac M, Dubeau F, Gotman J, 2005
  Seizure anticipation: do mathematical measures correlate with video-{EEG}
  evaluation?
\newblock \emph{Epilepsia} \textbf{46}, 385--396

\bibitem{Federico2005}
Federico P, Abbott DF, Briellmann RS, Harvey AS, Jackson GD, 2005 Functional
  {MRI} of the pre-ictal state.
\newblock \emph{Brain} \textbf{128}, 1811--1817

\bibitem{Meier2007}
Meier R, H\"aussler U, Aertsen A, Deransart C, Depaulis A, Egert U, 2007
  Short-term changes in bilateral hippocampal coherence precede epileptiform
  events.
\newblock \emph{NeuroImage} \textbf{38}, 138--149

\bibitem{Feldwisch2011}
Feldwisch-Drentrup H, Ihle M, {Le Van Quyen} M, Teixeira C, Dourado A, Timmer
  J, Sales F, Navarro V, Schulze-Bonhage A, Schelter B, 2011 Anticipating the
  unobserved: Prediction of subclinical seizures.
\newblock \emph{Epilepsy Behav.} \textbf{22}, {S119--S126}.
\newblock \doi{10.1016/j.yebeh.2011.08.023}

\bibitem{Aarabi2012}
Aarabi A, He B, 2012 A rule-based seizure prediction method for focal
  neocortical epilepsy.
\newblock \emph{Clin. Neurophysiol.} \textbf{123}, 1111--1122.
\newblock \doi{10.1016/j.clinph.2012.01.014}

\bibitem{Seyal2014}
Seyal M, 2014 Frontal hemodynamic changes precede {EEG} onset of temporal lobe
  seizures.
\newblock \emph{Clin. Neurophysiol.} \textbf{125}, 442--448.
\newblock \doi{10.1016/j.clinph.2013.09.003}

\bibitem{Arnhold1999}
Arnhold J, Grassberger P, Lehnertz K, Elger CE, 1999 A robust method for
  detecting interdependences: application to intracranially recorded {EEG}.
\newblock \emph{Physica~D} \textbf{134}, 419--430

\bibitem{Osterhage2007}
Osterhage H, Mormann F, Staniek M, Lehnertz K, 2007 Measuring synchronization
  in the epileptic brain: A comparison of different approaches.
\newblock \emph{Int. J. Bifurcation Chaos Appl. Sci. Eng.} \textbf{17},
  3539--3544

\bibitem{Osterhage2007b}
Osterhage H, Mormann F, Wagner T, Lehnertz K, 2007 Measuring the directionality
  of coupling: phase versus state space dynamics and application to {EEG} time
  series.
\newblock \emph{Int. J. Neural. Syst.} \textbf{17}, 139--148

\bibitem{Osterhage2008}
Osterhage H, Mormann F, Wagner T, Lehnertz K, 2008 Detecting directional
  coupling in the human epileptic brain: Limitations and potential pitfalls.
\newblock \emph{Phys. Rev.~E} \textbf{77}, 011914

\bibitem{Prusseit2008a}
Prusseit J, Lehnertz K, 2008 Measuring interdependences in dissipative
  dynamical systems with estimated {F}okker-{P}lanck coefficients.
\newblock \emph{Phys. Rev.~E} \textbf{77}, 041914

\bibitem{Wendling2009}
Wendling F, Bartolomei F, Senhadji L, 2009 Spatial analysis of intracerebral
  electroencephalographic signals in the time and frequency domain:
  identification of epileptogenic networks in partial epilepsy.
\newblock \emph{Phil. Trans. Roy. Soc. A} \textbf{367}, 297--316.
\newblock \doi{10.1098/rsta.2008.0220}

\bibitem{Andrzejak2011}
Andrzejak RG, Chicharro D, Lehnertz K, Mormann F, 2011 Using bivariate signal
  analysis to characterize the epileptic focus: The benefit of surrogates.
\newblock \emph{Phys. Rev.~E} \textbf{83}, 046203.
\newblock \doi{10.1103/PhysRevE.83.046203}

\bibitem{Andrzejak2011b}
Andrzejak RG, Kreuz T, 2011 Characterizing unidirectional couplings between
  point processes and flows.
\newblock \emph{{EPL}} \textbf{96}.
\newblock \doi{10.1209/0295-5075/96/50012}

\bibitem{Pikovsky2001}
Pikovsky AS, Rosenblum MG, Kurths J, 2001 \emph{Synchronization: {A} universal
  concept in nonlinear sciences}.
\newblock Cambridge, UK: Cambridge University Press.
\newblock \doi{10.1017/CBO9780511755743}

\bibitem{Boccaletti2002}
Boccaletti S, Kurths J, Osipov G, Valladares DL, Zhou CS, 2002 The
  synchronization of chaotic systems.
\newblock \emph{Phys. Rep.} \textbf{366}, 1--101.
\newblock \doi{10.1016/S0370-1573(02)00137-0}

\bibitem{Pereda2005}
Pereda E, Quian~Quiroga R, Bhattacharya J, 2005 Nonlinear multivariate analysis
  of neurophysiological signals.
\newblock \emph{Prog. Neurobiol.} \textbf{77}, 1--37.
\newblock \doi{10.1016/j.pneurobio.2005.10.003}

\bibitem{Gourevitch2006}
Gourevitch B, {Le Bouquin-Jeannes} R, Faucon G, 2006 Linear and nonlinear
  causality between signals: methods, examples and neurophysiological
  applications.
\newblock \emph{Biol. Cybern.} \textbf{95}, 349--369.
\newblock \doi{10.1007/s00422-006-0098-0}

\bibitem{Marwan2007}
Marwan N, Romano MC, Thiel M, Kurths J, 2007 Recurrence plots for the analysis
  of complex systems.
\newblock \emph{Phys. Rep.} \textbf{438}, 237--329.
\newblock \doi{10.1016/j.physrep.2006.11.001}

\bibitem{Lehnertz2009b}
Lehnertz K, Bialonski S, Horstmann MT, Krug D, Rothkegel A, Staniek M, Wagner
  T, 2009 Synchronization phenomena in human epileptic brain networks.
\newblock \emph{J.~Neurosci. Methods} \textbf{183}, 42--48.
\newblock \doi{10.1016/j.jneumeth.2009.05.015}

\bibitem{Friedrich2011}
Friedrich R, Peinke J, Sahimi M, Tabar MRR, 2011 Approaching complexity by
  stochastic methods: {F}rom biological systems to turbulence.
\newblock \emph{Phys. Rep.} \textbf{506}, 87--162.
\newblock \doi{10.1016/j.physrep.2011.05.003}

\bibitem{Lehnertz2011}
Lehnertz K, 2011 Assessing directed interactions from neurophysiological
  signals---an overview.
\newblock \emph{Physiol. Meas.} \textbf{32}, 1715--1724.
\newblock \doi{10.1088/0967-3334/32/11/R01}

\bibitem{Hlavackova2007}
Hlav\'a\v{c}kov\'a-Schindler K, Palu\v{s} M, Vejmelka M, Bhattacharya J, 2007
  Causality detection based on information-theoretic approaches in time series
  analysis.
\newblock \emph{Phys. Rep.} \textbf{441}, 1--46.
\newblock \doi{10.1016/j.physrep.2006.12.004}

\bibitem{Hao1989}
Hao BL, 1989 \emph{Elementary {S}ymbolic {D}ynamics and {C}haos in
  {D}issipative {S}ystems.}
\newblock Singapore: World Scientific

\bibitem{Bandt2002}
Bandt C, Pompe B, 2002 Permutation entropy: {A} natural complexity measure for
  time series.
\newblock \emph{Phys. Rev. Lett.} \textbf{88}, 174102

\bibitem{Daw2003}
Daw C, Finney C, Tracy E, 2003 A review of symbolic analysis of experimental
  data.
\newblock \emph{Rev. Sci. Instrum.} \textbf{74}, 915--930

\bibitem{Liu2004}
Liu Z, 2004 Measuring the degree of synchronization from time series data.
\newblock \emph{Europhys. Lett.} \textbf{68}, 19--25

\bibitem{Staniek2008}
Staniek M, Lehnertz K, 2008 Symbolic transfer entropy.
\newblock \emph{Phys. Rev. Lett.} \textbf{100}, 158101

\bibitem{Staniek2009}
Staniek M, Lehnertz K, 2009 Symbolic transfer entropy: inferring directionality
  in biosignals.
\newblock \emph{Biomed. Tech.} \textbf{54}, 323--328

\bibitem{Rulkov1995}
Rulkov NF, Sushchik MM, Tsimring LS, Abarbanel HDI, 1995 Generalized
  synchronization of chaos in directionally coupled chaotic systems.
\newblock \emph{Phys. Rev.~E} \textbf{51}, 980--994

\bibitem{Schreiber2000}
Schreiber T, 2000 Measuring information transfer.
\newblock \emph{Phys. Rev. Lett.} \textbf{85}, 461--464

\bibitem{Kaiser2002}
Kaiser A, Schreiber T, 2002 Information transfer in continuous processes.
\newblock \emph{Physica~D} \textbf{166}, 43--62

\bibitem{Verdes2005}
Verdes PF, 2005 Assessing causality from multivariate time series.
\newblock \emph{Phys. Rev.~E} \textbf{72}, 026222

\bibitem{Lungarella2007}
Lungarella M, Pitti A, Kuniyoshi Y, 2007 Information transfer at multiple
  scales.
\newblock \emph{Phys. Rev.~E} \textbf{76}, 056117.
\newblock \doi{10.1103/PhysRevE.76.056117}

\bibitem{Kowalski2010}
Kowalski AM, Martin MT, Plastino A, Zunino L, 2010 Information flow during the
  quantum-classical transition.
\newblock \emph{Phys. Lett. A} \textbf{374}, 1819--1826.
\newblock \doi{10.1016/j.physleta.2010.02.037}

\bibitem{Melzer2014}
Melzer A, Schella A, 2014 Symbolic transfer entropy analysis of the dust
  interaction in the presence of wakefields in dusty plasmas.
\newblock \emph{Phys. Rev. E} \textbf{89}, 041103.
\newblock \doi{10.1103/PhysRevE.89.041103}

\bibitem{Nian-Qiang2012}
Nian-Qiang L, Wei P, {Lian-Shan} Y, Bin L, {Ming-Feng} X, {Yi-Long} T, 2012
  Quantifying information flow between two chaotic semiconductor lasers using
  symbolic transfer entropy.
\newblock \emph{Chin. Phys. Lett.} \textbf{29}, 030502.
\newblock \doi{10.1088/0256-307X/29/3/030502}

\bibitem{Blain-Moraes2013}
{Blain-Moraes} S, Mashour GA, Lee H, Huggins JE, Lee U, 2013 Altered cortical
  communication in amyotrophic lateral sclerosis.
\newblock \emph{Neurosci. Lett.} \textbf{543}, 172--176.
\newblock \doi{10.1016/j.neulet.2013.03.028}

\bibitem{Jun2012}
Jun W, {Zheng-Feng} Y, 2012 Symbolic transfer entropy-based premature signal
  analysis.
\newblock \emph{Chin. Phys. B} \textbf{21}, 018702.
\newblock \doi{10.1088/1674-1056/21/1/018702}

\bibitem{Ku2011}
Ku SW, Lee U, Noh GJ, Jun IG, Mashour GA, 2011 Preferential inhibition of
  frontal-to-parietal feedback connectivity is a neurophysiologic correlate of
  general anesthesia in surgical patients.
\newblock \emph{PLoS One} \textbf{6}, e25155.
\newblock \doi{10.1371/journal.pone.0025155}

\bibitem{Jordan2013}
Jordan D, \emph{et~al.}, 2013 Simultaneous electroencephalographic and
  functional magnetic resonance imaging indicate impaired cortical top-down
  processing in association with anesthetic-induced unconsciousness.
\newblock \emph{Anesthesiology} \textbf{119}, 1031--1042.
\newblock \doi{10.1097/ALN.0b013e3182a7ca92}

\bibitem{Lee2013}
Lee U, Ku SW, Noh GJ, Baek SH, Choi BM, Mashour GA, 2013 Disruption of
  frontal-parietal communication by ketamine, propofol, and sevoflurane.
\newblock \emph{Anesthesiology} \textbf{118}, 1264--1275.
\newblock \doi{10.1097/ALN.0b013e31829103f5}

\bibitem{Untergehrer2014}
Untergehrer G, Jordan D, Kochs EF, Ilg R, Schneider G, 2014 Fronto-parietal
  connectivity is a non-static phenomenon with characteristic changes during
  unconsciousness.
\newblock \emph{PLoS One} \textbf{9}, e87498.
\newblock \doi{10.1371/journal.pone.0087498}

\bibitem{Martini2011}
Martini M, Kranz TA, Wagner T, Lehnertz K, 2011 Inferring directional
  interactions from transient signals with symbolic transfer entropy.
\newblock \emph{Phys. Rev.~E} \textbf{83}, 011919.
\newblock \doi{10.1103/PhysRevE.83.011919}

\bibitem{Kugiumtzis2013b}
Kugiumtzis D, 2013 Partial transfer entropy on rank vectors.
\newblock \emph{Eur. Phys. J.-Spec. Top.} \textbf{222}, 401--420.
\newblock \doi{10.1140/epjst/e2013-01849-4}

\bibitem{Papana2013}
Papana A, Kyrtsou C, Kugiumtzis D, Diks C, 2013 Simulation study of direct
  causality measures in multivariate time series.
\newblock \emph{Entropy} \textbf{15}, 2635--2661.
\newblock \doi{10.3390/e15072635}

\bibitem{Hahs2011}
Hahs DW, Pethel SD, 2011 Distinguishing anticipation from causality:
  Anticipatory bias in the estimation of information flow.
\newblock \emph{Phys. Rev. Lett.} \textbf{107}, 128701.
\newblock \doi{10.1103/PhysRevLett.107.128701}

\bibitem{Barnett2012}
Barnett L, Bossomaier T, 2012 Transfer entropy as a log-likelihood ratio.
\newblock \emph{Phys. Rev. Lett.} \textbf{109}, 138105.
\newblock \doi{10.1103/PhysRevLett.109.138105}

\bibitem{Smirnov2013}
Smirnov DA, 2013 Spurious causalities with transfer entropy.
\newblock \emph{Phys. Rev. E} \textbf{87}, 042917.
\newblock \doi{10.1103/PhysRevE.87.042917}

\bibitem{Haruna2013}
Haruna T, Nakajima K, 2013 Permutation complexity and coupling measures in
  hidden markov models.
\newblock \emph{Entropy} \textbf{15}, 3910--3930.
\newblock \doi{10.3390/e15093910}

\bibitem{Palus2007}
Palu\v{s} M, Vejmelka M, 2007 Directionality of coupling from bivariate time
  series: How to avoid false causalities and missed connections.
\newblock \emph{Phys. Rev.~E} \textbf{75}, 056211

\bibitem{Waddell2007}
Waddell J, Dzakpasu R, Booth V, Riley B, Reasor J, Poe G, Zochowski M, 2007
  Causal entropies--{A} measure for determining changes in the temporal
  organization of neural systems.
\newblock \emph{J.~Neurosci. Methods} \textbf{162}, 320--332

\bibitem{Engel1993b}
{Engel Jr} J, van Ness PC, Rasmussen TB, Ojemann LM, 1993 Outcome with respect
  to epileptic seizures.
\newblock In J~{Engel Jr}, ed., \emph{Surgical Treatment of the Epilepsies},
  609. New York: Raven Press

\bibitem{Blanco1995}
Blanco S, Garcia H, Quian~Quiroga R, Romanelli L, Rosso OA, 1995 Stationarity
  of the {EEG} series.
\newblock \emph{IEEE Eng. Med. Biol.} \textbf{4}, 395--399.
\newblock \doi{10.1109/51.395321}

\bibitem{Cao2004}
Cao Y, Tung W, Gao JB, Protopopescu VA, Hively LM, 2004 Detecting dynamical
  changes in time series using the permutationentropy.
\newblock \emph{Phys. Rev.~E} \textbf{70}, 046217

\bibitem{Staniek2007}
Staniek M, Lehnertz K, 2007 Parameter selection in permutation entropy
  measurements.
\newblock \emph{Int. J. Bifurcation Chaos Appl. Sci. Eng.} \textbf{17}, 3729

\bibitem{Palus2001a}
Palu\v{s} M, Kom\'arek V, Hrn\v{c}\'i\v{r} Z, \v{S}t\v{e}rbov\'a K, 2001
  Synchronization and information flow in {EEG}s of epileptic patients.
\newblock \emph{Med. Biol. Mag.} \textbf{20}, 65--71

\bibitem{Osterhage2008a}
Osterhage H, Bialonski S, Staniek M, Schindler K, Wagner T, Elger CE, Lehnertz
  K, 2008 Bivariate and multivariate time series analysis techniques and their
  potential impact for seizure prediction.
\newblock 189--208. Berlin: Wiley-VCH

\bibitem{BenJacob2007a}
{Ben-Jacob} E, Boccaletti S, Pomyalov A, Procaccia I, Towle VL, 2007 Detecting
  and localizing the foci in human epileptic seizures.
\newblock \emph{Chaos} \textbf{17}, 043113

\bibitem{Lai2007}
Lai YC, Frei MG, Osorio I, Huang L, 2007 Characterization of synchrony with
  applications to epileptic brain signals.
\newblock \emph{Phys. Rev. Lett.} \textbf{98}, 108102

\bibitem{Schevon2007}
Schevon CA, \emph{et~al.}, 2007 Cortical abnormalities in epilepsy revealed by
  local {EEG} synchrony.
\newblock \emph{NeuroImage} \textbf{35}, 140--148

\bibitem{Ortega2008}
Ortega GJ, {Menendez de la Prida} L, Sola RG, Pastor J, 2008 Synchronization
  clusters of interictal activity in the lateral temporal cortex of epileptic
  patients: {I}ntraoperative electrocorticographic analysis.
\newblock \emph{Epilepsia} \textbf{49}, 269--280

\bibitem{Zaveri2009}
Zaveri HP, Pincus SM, Goncharova II, Duckrow RB, Spencer DD, Spencer SS, 2009
  Localization-related epilepsy exhibits significant connectivity away from the
  seizure-onset area.
\newblock \emph{NeuroReport} \textbf{20}, 891--895

\bibitem{Warren2010}
Warren C, Hu S, Stead M, Brinkmann BH, Bower MR, Worrell GA, 2010 Synchrony in
  normal and focal epileptic brain: {T}he seizure onset zone is functionally
  disconnected.
\newblock \emph{J.~Neurophysiol.} \textbf{104}, 3530--3539

\bibitem{Bettus2011}
Bettus G, \emph{et~al.}, 2011 Interictal functional connectivity of human
  epileptic networks assessed by intracerebral {EEG} and {BOLD} signal
  fluctuations.
\newblock \emph{PLoS ONE} \textbf{6}, e20071.
\newblock \doi{10.1371/journal.pone.0020071}

\bibitem{Ortega2011}
Ortega GJ, Peco IH, Sola RG, Pastor J, 2011 Impaired mesial synchronization in
  temporal lobe epilepsy.
\newblock \emph{Clin. Neurophysiol.} \textbf{122}, 1106--1116.
\newblock \doi{10.1016/j.clinph.2010.11.001}

\bibitem{Palmigiano2012}
Palmigiano A, Pastor J, Garcia~de Sola R, Ortega GJ, 2012 Stability of
  synchronization clusters and seizurability in temporal lobe epilepsy.
\newblock \emph{PLoS ONE} \textbf{7}, e41799.
\newblock \doi{10.1371/journal.pone.0041799}

\bibitem{Ansari2006}
Ansari-Asl K, Senhadji L, Bellanger JJ, Wendling F, 2006 Quantitative
  evaluation of linear and nonlinear methods characterizing interdependencies
  between brain signals.
\newblock \emph{Phys. Rev.~E} \textbf{74}, 031916

\bibitem{Lehnertz2011b}
Lehnertz K, Krug D, Staniek M, Gl\"usenkamp D, Elger CE, 2011 Preictal directed
  interactions in epileptic brain networks.
\newblock 265--272. Boca Raton, FL, USA: CRC Press

\bibitem{Mormann2008b}
Mormann F, Osterhage H, Andrzejak RG, Weber B, Fernandez G, Fell J, Elger CE,
  Lehnertz K, 2008 Independent delta/theta rhythms in the human hippocampus and
  entorhinal cortex.
\newblock \emph{Front. Hum. Neurosci.} \textbf{2}, 3.
\newblock \doi{10.3389/neuro.09.003.2008}

\bibitem{Mormann2000}
Mormann F, Lehnertz K, David P, Elger CE, 2000 Mean phase coherence as a
  measure for phase synchronization and its application to the {EEG} of
  epilepsy patients.
\newblock \emph{Physica~D} \textbf{144}, 358--369.
\newblock \doi{10.1016/S0167-2789(00)00087-7}

\bibitem{Mormann2003a}
Mormann F, Andrzejak R, Kreuz T, Rieke C, David P, Elger CE, Lehnertz K, 2003
  Automated detection of a preseizure state based on a decrease in
  synchronization in intracranial electroencephalogram recordings from epilepsy
  patients.
\newblock \emph{Phys. Rev.~E} \textbf{67}, 021912.
\newblock \doi{10.1103/PhysRevE.67.021912}

\bibitem{Winterhalder2006c}
Winterhalder M, Schelter B, Maiwald T, Brandt A, Schad A, {Schulze-Bonhage} A,
  Timmer J, 2006 Spatio-temporal patient-individual assessment of
  synchronization changes for epileptic seizure prediction.
\newblock \emph{Clin. Neurophysiol.} \textbf{117}, 2399--2413

\bibitem{Kuhlmann2010}
Kuhlmann L, Freestone D, Lai AL, Burkitt AN, Fuller K, Grayden D, Seiderer L,
  Vogrin S, Mareels IMY, Cook MJ, 2010 Patient-specific
  bivariate-synchrony-based seizure prediction for short prediction horizons.
\newblock \emph{Epilepsy Res.} \textbf{91}, 214--231.
\newblock \doi{10.1016/j.eplepsyres.2010.07.014}

\bibitem{Wang2011b}
Wang L, Wang C, Fu F, Yu X, Guo H, C CX, Jing X, Zhang H, Dong X, 2011 Temporal
  lobe seizure prediction based on a complex {G}aussian wavelet.
\newblock \emph{Clin. Neurophysiol.} \textbf{122}, 656--663.
\newblock \doi{10.1016/j.clinph.2010.09.018}

\bibitem{Zheng2014}
Zheng Y, Wang G, Li K, Bao G, Wang J, 2014 Epileptic seizure prediction using
  phase synchronization based on bivariate empirical mode decomposition.
\newblock \emph{Clin. Neurophysiol.} \textbf{125}, 1104--1111.
\newblock \doi{10.1016/j.clinph.2013.09.047}

\bibitem{Prince1967}
Prince DA, Wilder J, 1967 Control mechanisms in cortical epileptogenic foci.
  "surround" inhibition.
\newblock \emph{Arch. Neurol.} \textbf{16}, 194--202

\bibitem{Elger1983}
Elger CE, Speckmann EJ, 1983 Penicillin induced epileptic foci in the motor
  cortex: vertical inhibition.
\newblock \emph{Electroencephalogr. Clin. Neurophysiol.} \textbf{56}, 604--622

\bibitem{Andrzejak2003}
Andrzejak RG, Mormann F, Kreuz T, Rieke C, Kraskov A, Elger CE, Lehnertz K,
  2003 Testing the null hypothesis of the nonexistence of a preseizure state.
\newblock \emph{Phys. Rev.~E} \textbf{67}, 010901(R)

\bibitem{Andrzejak2009}
Andrzejak RG, Chicharro D, Elger CE, Mormann F, 2009 Seizure prediction: {A}ny
  better than chance?
\newblock \emph{Clin. Neurophysiol.} \textbf{120}, 1465--1478.
\newblock \doi{10.1016/j.clinph.2009.05.019}

\bibitem{Winterhalder2003}
Winterhalder M, Maiwald T, Voss HU, Aschenbrenner-Scheibe R, Timmer J,
  Schulze-Bonhage A, 2003 The seizure prediction characteristic: A general
  framework to assess and compare seizure prediction methods.
\newblock \emph{Epilepsy Behav.} \textbf{3}, 318--325

\bibitem{Kreuz2004}
Kreuz T, Andrzejak RG, Mormann F, Kraskov A, St\"ogbauer H, Elger CE, Lehnertz
  K, Grassberger P, 2004 Measure profile surrogates: A method to validate the
  performance of epileptic seizure prediction algorithms.
\newblock \emph{Phys. Rev.~E} \textbf{69}, 061915

\bibitem{Schelter2006}
Schelter B, Winterhalder M, Maiwald T, Brandt A, Schad A, Schulze-Bonhage A,
  Timmer J, 2006 Testing statistical significance of multivariate time series
  analysis techniques for epileptic seizure prediction.
\newblock \emph{Chaos} \textbf{16}, 013108

\bibitem{Wong2007}
Wong S, Gardner AB, Krieger AM, Litt B, 2007 A stochastic framework for
  evaluating seizure prediction algorithms using hidden {M}arkov models.
\newblock \emph{J.~Neurophysiol.} \textbf{97}, 2525--2532

\bibitem{Snyder2008}
Snyder DE, Echauz J, Grimes DB, Litt B, 2008 The statistics of a practical
  seizure warning system.
\newblock \emph{J.~Neural. Eng.} \textbf{5}, 392--401

\bibitem{Porta2002}
Porta A, Furlan R, Rimoldi O, Pagani M, Malliani A, van~de Borne P, 2002
  Quantifying the strength of the linear causal coupling in closed loop
  interacting cardiovascular variability signals.
\newblock \emph{Biol. Cybern.} \textbf{86}, 241--251.
\newblock \doi{10.1007/s00422-001-0292-z}

\bibitem{Thiel2006}
Thiel M, Romano MC, Kurths J, Rolfs M, Kliegl R, 2006 Twin surrogates to test
  for complex synchronisation.
\newblock \emph{Europhys. Lett.} \textbf{75}, 535--541

\bibitem{Romano2009}
Romano MC, Thiel M, Kurths J, Mergenthaler K, Engbert R, 2009 Hypothesis test
  for synchronization: Twin surrogates revisited.
\newblock \emph{Chaos} \textbf{19}, 015108.
\newblock \doi{10.1063/1.3072784}

\bibitem{Faes2010}
Faes L, Porta A, Nollo G, 2010 Testing frequency-domain causality in
  multivariate time series.
\newblock \emph{IEEE Trans. Biomed. Eng.} \textbf{57}, 1897--1906.
\newblock \doi{10.1109/TBME.2010.2042715}

\bibitem{Franaszczuk1994}
Franaszczuk PJ, Bergey GK, Kaminski MJ, 1994 Analysis of mesial temporal
  seizure onset and propagation using the directed transfer function method.
\newblock \emph{Electroencephalogr. Clin. Neurophysiol.} \textbf{91}, 413--427

\bibitem{Franaszczuk1998}
Franaszczuk PJ, Bergey GK, 1998 Application of the directed transfer function
  method to mesial and lateral onset temporal lobe seizures.
\newblock \emph{Brain Topogr.} \textbf{1}, 13--21

\bibitem{LeVanQuyen1998}
{Le Van Quyen} M, Adam C, Baulac M, Martinerie J, Varela FJ, 1998 Nonlinear
  interdependencies of {EEG} signals in human intracranially recorded temporal
  lobe seizures.
\newblock \emph{Brain Res.} \textbf{792}, 24

\bibitem{Wendling2001}
Wendling F, Bartolomei F, Bellanger J, Chauvel P, 2001 Interpretations of
  interdependencies in epileptic signals using a macroscopic physiological
  model of the {EEG}.
\newblock \emph{Clin. Neurophysiol.} \textbf{112}, 1201--1218

\bibitem{Chavez2003}
Chavez M, Martinerie J, {Le Van Quyen} M, 2003 Statistical assessment of
  nonlinear causality: application to epileptic {EEG} signals.
\newblock \emph{J.~Neurosci. Methods} \textbf{124}, 113--128

\bibitem{Smirnov2005b}
Smirnov DA, Bodrov MB, Velazquez JLP, Wennberg RA, Bezruchko BP, 2005
  Estimation of coupling between oscillators from short time series via phase
  dynamics modeling: Limitations and application to {EEG} data.
\newblock \emph{Chaos} \textbf{15}, 024102

\bibitem{Wilke2009}
Wilke C, van Drongelen W, Kohrman M, He B, 2009 Identification of epileptogenic
  foci from causal analysis of {ECoG} interictal spike activity.
\newblock \emph{Clin. Neurophysiol.} \textbf{120}, 1449--1456.
\newblock \doi{10.1016/j.clinph.2009.04.024}

\bibitem{Nunez1995}
Nunez PL, 1995 \emph{Neocortical Dynamics and Human {EEG} Rhythms}.
\newblock Oxford, UK: Oxford University Press

\bibitem{Frenzel2007}
Frenzel S, Pompe B, 2007 Partial mutual information for coupling analysis of
  multivariate time series.
\newblock \emph{Phys. Rev. Lett.} \textbf{99}, 204101.
\newblock \doi{10.1103/PhysRevLett.99.204101}

\bibitem{Pompe2011}
Pompe B, Runge J, 2011 Momentary information transfer as a coupling measure of
  time series.
\newblock \emph{Phys. Rev.~E} \textbf{83}, 051122.
\newblock \doi{10.1103/PhysRevE.83.051122}

\bibitem{Runge2012}
Runge J, Heitzig J, Petoukhov V, Kurths J, 2012 Escaping the curse of
  dimensionality in estimating multivariate transfer entropy.
\newblock \emph{Phys. Rev. Lett.} \textbf{108}, 258701.
\newblock \doi{10.1103/PhysRevLett.108.258701}

\bibitem{Runge2012b}
Runge J, Heitzig J, Marwan N, Kurths J, 2012 Quantifying causal coupling
  strength: A lag-specific measure for multivariate time series related to
  transfer entropy.
\newblock \emph{Phys. Rev. E} \textbf{86}, 061121.
\newblock \doi{10.1103/PhysRevE.86.061121}

\bibitem{Dickten2014}
Dickten H, Lehnertz K, 2014 Identifying delayed directional couplings with
  symbolic transfer entropy.
\newblock \emph{Phys. Rev. E} \textbf{90}, 062706.
\newblock \doi{10.1103/PhysRevE.90.062706}

\end{thebibliography}
\end{document}